\documentclass[aip,jcp,amsmath,amssymb,preprint,numeric]{revtex4-1}
\usepackage{graphicx}
\usepackage{subfigure}
\usepackage{color}

\begin{document}
\title {Unraveling the success and failure of mode coupling theory from consideration of entropy}
\author{Manoj Kumar Nandi}
\thanks{M. K. Nandi and A. Banerjee contributed equally to this work.}
\affiliation{\textit{Polymer Science and Engineering Division, CSIR-National Chemical Laboratory, Pune-411008, India}}
\author{Atreyee Banerjee}
\thanks{M. K. Nandi and A. Banerjee contributed equally to this work.}
\affiliation{\textit{Polymer Science and Engineering Division, CSIR-National Chemical Laboratory, Pune-411008, India}}
\author{Shiladitya Sengupta}
\affiliation{\textit{Department of Chemical Physics, Weizmann Institute of Science, Rehovot 76100, Israel}}
\author{ Srikanth Sastry}
\affiliation{\textit{Theoretical Sciences Unit, Jawaharlal Nehru Centre for Advanced Scientific Research, Jakkur Campus, Bengaluru 560 064, India}}
\author{Sarika Maitra Bhattacharyya}
\email{mb.sarika@ncl.res.in}
\affiliation{\textit{Polymer Science and Engineering Division, CSIR-National Chemical Laboratory, Pune-411008, India}}

\date{\today}

\begin{abstract}

We analyze the dynamics of model supercooled liquids in a temperature regime where predictions of mode coupling theory (MCT) are known to be valid qualitatively.
 In this regime, the Adam-Gibbs (AG) relation, based on an activation picture of dynamics also describes the dynamics satisfactorily, and we explore the mutual 
consistency and interrelation of these descriptions. Although entropy and dynamics are related via phenomenological theories, the connection between MCT
  and entropy has not been argued for. In this work we explore this connection and provide a microscopic derivation of the phenomenological Rosenfeld theory.
  At low temperatures the overlap between MCT power law regime and AG relation implies that the AG relation predicts an avoided divergence at $T_c$, 
the origin of which is traced back to the vanishing of pair configurational entropy, which we find occurs at the same temperature. We also show that 
 the residual multiparticle entropy plays an important role in describing the relaxation time. 

\end{abstract}
\maketitle
\section{Introduction}
In the study of liquid state physics, the structure of the liquid, which is  often described primarily by the two body radial distribution function (rdf),
 has always played a central role.  The structure can not only describe the thermodynamic properties of the liquid like the energy and pressure, under certain theoretical 
frameworks like the mode coupling theory the structure can also determine the dynamics \cite{sjogren,Gotze}.  In a series of paper Berthier and  Tarjus have described
 the behaviour of  two systems with different interaction potentials, namely, the Lennard-Jones (LJ) and the Weeks-Chandler-Andersen (WCA)
 potentials. Although at the same temperature and density the structures of these systems  are very close, the dynamics display significant differences at low
 temperatures
 \cite{tarjus_prl,tarjus_pre,tarjus_epje,tarjus_berthier_jcp}. These studies questioned the role of structure in determining the dynamics.
 Coslovich has shown that although the two body radial distribution 
function of these two systems are quite similar, triplet correlations are significantly different \cite{coslovich-jcp}. He has also shown that the LJ system has more pronounced 
local ordering \cite{coslovich}. In supercooled liquids these locally preferred structures are known to form correlated domains which have been argued to give rise to the slow dynamics \cite{tarjus}.
 An estimation of this length scale of the domains and its connection to the relaxation timescale is a topic of ongoing research \cite {Smarajit_pre, Paddy-jcp13}. 
One such study by Hocky {\it et al.} has shown that the point-to-set correlation length of the LJ system is larger compared to that of the WCA system and that this difference in 
correlation length can account for the difference in dynamics of the two systems \cite{hocky-prl}. From these studies one may conclude that the difference in dynamics 
primarily comes from many body correlations. However, in a recent study by some of us it has been shown that two body correlation information is good enough to capture the difference in
 the dynamics between the two systems. The study also reveals that the divergence temperature at which an approximation to the configurational entropy using pair correlation alone goes to zero,
 is similar  to the
mode coupling theory (MCT) transition temperature, $T_c$ \cite{bssb}. As mentioned before, MCT is a microscopic theory where the structural inputs determine the dynamics.
Although entropy and dynamics are related {\it via} phenomenological 
Rosenfeld \cite{Rosenfeld} and Adam-Gibbs (AG) \cite{adam-gibbs} relations at high and low temperatures respectively,  MCT does not have any apparent 
connection to entropy. Thus it is of great interest to try to understand the origin of the coincidence of the MCT divergence temperature and the temperature where pair
configurational entropy goes to zero.

At normal liquid temperatures, a semi quantitative correlation between the dynamics (transport properties) 
and thermodynamics (excess entropy), proposed by Rosenfeld \cite {Rosenfeld,Rosenfeld-iop}, has been extensively studied in recent times,
 where the relaxation time $\tau$ can be written as,
\begin{equation}
 \tau(T)=C \exp \left[ -KS_{ex}\right]
\label{rosenfeld}
\end{equation}

\noindent Here $C$ and  $K$ are the constants. Since the pair entropy $S_{2}$, which is obtained only from the pair correlation function, accounts for $80\%-90\%$ of the excess entropy
 \cite {Baranyai-cp,bssb}, many simulation studies have replaced $S_{ex}$ by $S_2$ and have shown that even with $S_2$ the transport
coefficients follow Rosenfeld scaling \cite{tarjus_berthier_jcp,Dzugutov,hoyt, trusket_2006,Ruchi_charu_2006}.

Bagchi and coworkers used Zwanzig's rugged energy landscape model of diffusion \cite{zwanzig} and by connecting the ruggedness to 
the excess entropy have provided a derivation of Rosenfeld relation \cite{saikat_jcp}.
 Samanta {\it et al} \cite {alok} have shown that under certain approximations the diffusion coefficient as obtained from MCT follows Rosenfeld scaling.
 Das and coworkers have performed microscopic MCT calculations which show that the diffusion values thus obtained can be fitted to Rosenfeld scaling \cite{shankar_das}.
 Some of these studies have reported that the scaling parameter is not unique, hence the whole temperature region cannot be fitted to a single straight
 line \cite{shankar_das,manish_charu}.


Although Rosenfeld scaling holds at high temperature, it is  known to breakdown at low temperatures even with multiple scaling parameters. 
At low temperatures the correlation
 between the transport coefficients and entropy is usually described by the well known Adam-Gibbs relation \cite{adam-gibbs}, 
\begin{equation}
\tau(T)=\tau_{o}\exp\left(\frac{A}{TS_{c}}\right),
\label{ag}
\end{equation}
where $S_c$ is the configurational entropy of the system. For a wide range of systems, the AG relation is found to hold \cite{bssb,shila-jcp,shila-prl} below a
 moderately high temperature referred to as the onset temperature of slow dynamics. 

In this paper, we explore the connection between dynamics, characterization of structure as contained in the pair and higher order correlations,
 and entropy, and relations between them as described by MCT and the AG relation, using computer simulations of two model liquids and analytical results that seek 
to relate descriptions of dynamics in terms of structure, and entropy. 
Our present study shows that the AG theory, which is based on activation dynamics can completely describe the mode coupling theory (MCT) power law behavior
 in the region where the latter is found to be valid.  An earlier study also observing similar overlap region \cite{sciortino-pre-2002}
 argued that the observation supports the hypothesis 
that a direct relation exists between the number of basins and their connectivity \cite{sciortino-prl-2000,sciortino-prl-2002}. In this work to understand the 
above mentioned observations,
we explore the connection between mode coupling theory (MCT) and entropy and discuss different predictions of MCT in the light of entropy. 
We also analyze the different roles of pair and many body correlations.

 Although MCT makes predictions about dynamics in both Rosenfeld and AG temperature regimes, no connection between  MCT and entropy has been argued for, 
except for one study \cite{alok}, as mentioned earlier.  We show that, under some assumptions, 
 the memory function in the MCT equation for structural relaxation is related to the pair excess entropy, thus providing a  microscopic derivation of the phenomenological
 Rosenfeld expression for the structural relaxation time, $\tau$. 
Our study also can explain the origin of the temperature dependence of the Rosenfeld parameter. The
 origin of higher relaxation time and higher activation energy as predicted by MCT is also obtained from the analysis of the memory function. 

   As mentioned above the AG expression for relaxation times and the MCT power law form overlap in a certain temperature regime.
The AG relation is valid for a wide temperature range which includes the range in which the MCT power law prediction holds.
Thus in the MCT regime, the relaxation time follows both the AG and 
MCT behaviour. Our study reveals that the origin of the avoided divergence like behaviour (as given by MCT power law) in the AG relation is related to the 
vanishing of the pair configurational entropy.  However we show that the pair configurational entropy, although predicting the correct MCT transition temperature,
 by itself cannot
predict the MCT power law behaviour. The residual multiparticle entropy (RMPE) plays an important role in providing the correct temperature dependence of relaxation times.
 We also find  a connection between the AG coefficient (A), pair thermodynamic fragility ($K_{T2}$) and  MCT critical exponent ($\gamma$).
 We  show that although both `A' and $K_{T2}$ are dependent on density,  their ratio which is related to $\gamma$ is density-independent.

The paper is organized as follows:
The simulation details are given in Sec. II. In Sec. III we describe
the methods used for evaluating the various quantities of interest and provide other necessary background. In Sec-IV we report some observations that motivate
 our analytical results which are  described in Sec-V. In Sec-VI we present additional numerical results and their analysis. Sec. VII contains a discussion of
 presented results and conclusions.




\section{Simulation Details}
We have performed molecular dynamics simulations of the Kob-Andersen model which is a binary mixture (80:20) of Lennard-Jones (LJ) particles and the 
corresponding WCA version \cite{kob,chandler}. 
 The interatomic pair  
potential between species {\it i} and {\it j}, with ${\it i,j}= A,B$, 
$U_{ij}(r)$ is described by a shifted and truncated Lennard-Jones (LJ) potential, as given by:
\begin{equation}
 U_{ij}(r)=
\begin{cases}
 U_{ij}^{(LJ)}(r;\sigma_{ij},\epsilon_{ij})- U_{ij}^{(LJ)}(r^{(c)}_{ij};\sigma_{ij},\epsilon_{ij}),    & r\leq r^{(c)}_{ij}\\
   0,                                                                                       & r> r^{(c)}_{ij}
\end{cases}
\end{equation}

\noindent where $U_{ij}^{(LJ)}(r;\sigma_{ij},\epsilon_{ij})=4\epsilon_{ij}[({\sigma_{ij}}/{r})^{12}-({\sigma_{ij}}/{r})^{6}]$ and
 $r^{(c)}_{ij}=2.5\sigma_{ij}$ for the LJ systems and $r^{(c)}_{ij}$  is equal to the position of the minimum of $U_{ij}^{(LJ)}$
for the WCA systems. Length, temperature and
time are given in units of $\sigma_{11}$, ${k_{B}T}/{\epsilon_{11}}$ and $\tau = \surd({m_1\sigma_{11}^2}/{\epsilon_{11}})$, 
respectively.  
Here we have simulated Kob Andersen Model  
with the interaction parameters  $\sigma_{11}$ = 1.0, $\sigma_{12}$ =0.8 ,$\sigma_{22}$ =0.88,  $\epsilon_{11}$ =1, $\epsilon_{12}$ =1.5,
 $\epsilon_{22}$ =0.5, $m_{1}$ = $m_2$=1.0 .

The molecular dynamics (MD) simulations have been carried out using the LAMMPS 
package \cite{lammps}.
We have performed MD simulations in the canonical ensemble (NVT) using  Nos\'{e}-Hoover thermostat  with integration timestep 0.005$\tau$. The time
constants for  Nos\'{e}-Hoover thermostat  are taken to be 100  timesteps.
The sample is kept in a cubic box with periodic boundary condition.
 System size is $N = 500$, $N_A = 400$ (N $=$ total number
of particles, $N_A$ $=$ number of particles of type A) and we have studied a broad range of density $\rho$ from 1.2 to
1.6. For all state points, three to five independent samples with run lengths $>$ 100$\tau_{\alpha}$ ($\tau_{\alpha}$ is the $\alpha$-
relaxation time) are analyzed.

\section{Definitions and Background}

\subsection{Relaxation time}
We have calculated the relaxation times from the decay of the
overlap function q(t), using $q(t = \tau_{\alpha} , T )/N =
1/e$. The overlap function is defined as
\begin{eqnarray}
\langle q(t) \rangle \equiv \left \langle \int dr \rho(r, t_0 )\rho(r, t + t_0 )\right \rangle \nonumber\\
=\left \langle \sum_{i=1}^{N}\sum_{j=1}^{N} \delta({\bf{r}}_j(t_0)-{\bf{r}}_i(t+t_0)) \right \rangle \nonumber\\
=\left \langle \sum_{i=1}^{N} \delta({\bf{r}}_i(t_0)-{\bf{r}}_i(t+t_0)) \right \rangle \nonumber\\
+\left \langle \sum_{i}\sum_{j\neq i} \delta({\bf{r}}_i(t_0)-{\bf{r}}_j(t+t_0)) \right \rangle
\end{eqnarray}
The overlap function is a two-point time correlation
function of local density $\rho(r, t)$. It has been used in
many recent studies of slow relaxation \cite{shila-jcp}.
 In this
work, we consider only the self-part of the total overlap
function (i.e. neglect the $i \neq j$ terms in the double
summation). This approximation has been shown to be a good approximation to the full overlap function. So, the self part of the overlap function can be written as,
\begin{eqnarray}
 \langle q(t) \rangle \approx \left \langle \sum_{i=1}^{N} \delta({\bf{r}}_i(t_0)-{\bf{r}}_i(t+t_0)) \right \rangle
\end{eqnarray}

The $\delta$ function is approximated by a window function $\omega(x)$ which defines the
condition of “overlap” between two particle positions
separated by a time interval t:
\begin{eqnarray}
 \langle q(t) \rangle \approx \left \langle \sum_{i=1}^{N} \omega (\mid{\bf{r}}_i(t_0)-{\bf{r}}_i(t+t_0)\mid) \right \rangle \nonumber\\
\omega(x) = 1, x \leq {\text{a implying “overlap”}} \nonumber\\
=0, \text{otherwise}
\end{eqnarray}

The time dependent overlap function thus depends on
the choice of the cut-off parameter a, which we choose
to be $0.3$. This parameter is chosen such that particle positions separated due to small amplitude vibrational motion are treated as the same, or that $a^2$ is
 comparable to
the value of the MSD in the plateau between the ballistic
and diffusive regimes.

  Relaxation times obtained from the decay of the self intermediate scattering function $F_s (k, t)$ using the 
definition $F_s (k, t = \tau_\alpha , T)$ = $1/e$ at $k\simeq 2\pi /r_{max}$, where $r_{max}$ is
the first maximum of the radial distribution function
. The self
intermediate scattering function is calculated from the
simulated trajectory as
\begin{equation}
 F_s(k,t)=\frac{1}{N}\left \langle \sum_{i=1}^{N} \exp(-i{\bf{k}}.({\bf{r}}_i(t)-{\bf{r}}_i(0))) \right \rangle
\end{equation}

Since relaxation times from $q(t)$ and $F_s (k, t)$
behave very similarly at low temperature we have used the time scale
obtained from $q(t)$. As $q(t)$ cannot be calculated from MCT, at high temperatures where we compare the simulation result with the relaxation time
 obtained analytically from MCT we have computed relaxation time using $F_s (k, t)$. 

\subsection{Static Structure Factor}

We measure the partial structure factor $S_{\alpha \beta}(k)$ which are needed as input for the MCT calculations. They are defined as
\begin{equation}
 S_{\alpha \beta}(k)=\frac{1}{\sqrt { N_{\alpha} N_{\beta}}} \sum_{i=1}^{N_{\alpha}} \sum_{j=1}^{N_{\beta}} \exp(-i{\bf{k}}.({\bf{r}}_i^{\alpha}-{\bf{r}}_j^{\beta}))
\label{sq-eqn}
\end{equation}

\subsection{ Mode coupling Theory}
Many properties of a glass forming liquids can be explained by the well known mode coupling theory of the glass transition (MCT). This
microscopic theory can give a qualitative description of dynamical properties (such as temperature dependence of relaxation time) if the static
structure of the liquid is known and many experiments and simulation results has shown that MCT predictions hold good in the temperature regime of initial slow down
 of dynamics
\cite{Gotze}.  
The equation for the intermediate scattering function $\phi(k,t)$ is given by
\begin{equation}
 \ddot{\phi}(k,t)+\Gamma \dot{\phi}(k,t)+\Omega_k^2 \phi(k,t)+\Omega_k^2\int dt' \mathcal{M}(t-t')\dot{\phi}(k,t')=0
\label{fkt-eqn}
\end{equation}
where $\Omega_k^2=\frac{k^2k_BT}{mS(k)}$ and memory function of $\phi(k,t)$ can be written as :
\begin{eqnarray}
\label{memory-fkt}
 \mathcal{M}(k,t)=\frac{1}{2\rho k^2}\int \frac{d\bf{q}}{(2\pi)^3}{V_{k}^{2}({\bf{q}},{\bf{k-q}})S(k)S(\mid{\bf{k-q}}\mid)S(q)\phi(q,t)\phi(\mid{\bf{k-q}}\mid,t)}
\end{eqnarray}
where ${\bf{k-q}}=\bf{p}$ and $ V_k({\bf{q}},{\bf{p}})=[\hat{\bf{k}}.{\bf{q}}\rho C(q) + \hat{\bf{k}}.{\bf{p}}\rho C(p)]$ .

For the self intermediate scattering function a similar equation may be written, as 
\begin{equation}
 \ddot{\phi_s}(k,t)+\Gamma \dot{\phi_s}(k,t)+\Omega_0^2 \phi_s(k,t)+\Omega_0^2\int dt' \mathcal{M}_s(t-t')\dot{\phi_s}(k,t')=0
\label{fskt-eqn}
\end{equation}
where $\Omega_0^2=\frac{k^2k_BT}{m}$ and  memory function $\mathcal{M}_s(k,t)$ can be written as
\begin{eqnarray}
\label{memory-fskt}
\mathcal{M}_s(k,t)=\frac{1}{\rho k^2}\times\int_0^{\infty}\frac{d{\bf{q}}}{(2\pi)^3}[\hat{\bf{k}}.{\bf{q}}]^2(\rho C(q))^2 S(q)\phi(q,t)\phi_s(p,t) 
\end{eqnarray}

We need the static structure factor to solve these equations, which is obtained from 
computer simulation, by Eq.\ref{sq-eqn}. The temperature dependence of the system enters in MCT through S(q) and
since we need very precise S(q)  near the MCT transition, we simulated S(q) at three temperatures around
the transition point and used them to
create the structure factors at intermediate temperatures by quadratic interpolation method as
described in \cite{kob-nauroth-pre}.  

To solve the Eq.\ref{fskt-eqn} we need $\phi(k,t)$ as an input, which can be taken by solving Eq.\ref{fkt-eqn}.
Using these expressions we have calculated the relaxation time, $\tau_{MCT}$  from the relaxation of $\phi(k,t)$
 at $1/e$.\\

\subsection{Configurational Entropy}

Configurational entropy, $S_c$ per particle, the measure of
the number of distinct local energy minima, is calculated \cite{srikanth_PRL}
by subtracting from the total entropy of the system the
vibrational component:
$S_c (T ) = S_{total} (T ) - S_{vib} (T )$ \cite{shila-jcp,Srikanth_nature}.
The total entropy of the liquid is obtained via thermodynamic
integration from the ideal gas limit. Vibrational entropy is calculated by making a harmonic approximation to the potential energy about a given local minimum.

\subsection{Pair Configurational Entropy}

To get an estimate of the configurational entropy as predicted by the pair correlation we rewrite $S_{c}$ in terms of the pair contribution to 
configurational entropy $S_{c2}$\cite{bssb},
\begin{equation}
S_{c}=S_{id}+S_{ex}-S_{vib}=S_{id}+S_{2}+\Delta S-S_{vib}=S_{c2}+\Delta S
\label{sc2}
\end{equation}
\noindent
Where $S_{c2}=S_{id}+S_{2}-S_{vib}$.  $S_{ex}$ can be expanded in an infinite series, 
$S_{ex}=S_{2}+S_{3}+.....=S_{2}+\Delta S$ using Kirkwood's factorization \cite{Kirkwood} of the N-particle distribution function \cite{green_jcp,raveche,Wallace}. 
 $S_{n}$ is the $``n"$ body contribution to the entropy. Thus the pair excess entropy is $S_{2}$ and the higher order contributions to excess entropy is given by 
the residual multiparticle entropy (RMPE), $\Delta S=S_{ex}-S_{2}$

\section{Observations}

As the liquid is supercooled, the Rosenfeld scaling, observed to be valid at normal temperatures, is known to break down 
\cite{manish_charu}. However, in this regime the Adam-Gibbs relation is found to hold \cite{adam-gibbs,shila-jcp}. The Adam Gibbs relation explains the 
behaviour of dynamical property like relaxation time using configurational entropy which is a thermodynamical property. So this relation connects thermodynamics
and dynamics for low temperature liquids.  In the Adam-Gibbs relation it is not the excess entropy but the configurational entropy which dictates the dynamics.

\begin{figure}[h]
\centering
\includegraphics[width=0.45\textwidth]{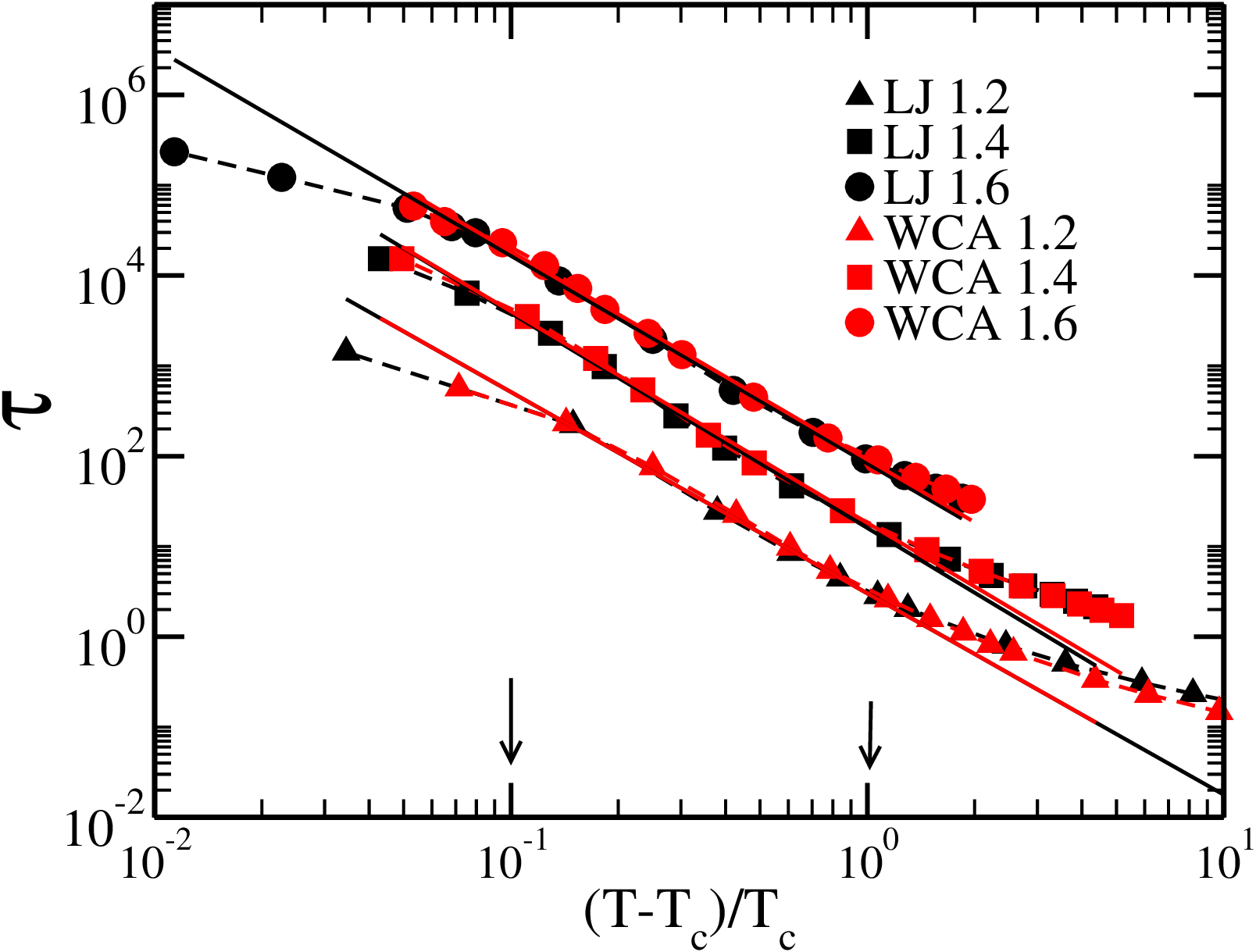}
\caption{\it{ The power law behaviour of relaxation time of numerical simulation, $\tau$, as predicted by MCT (Eq.\ref{t-tc})͒ appears as a straight line
for a certain region ($10^{-1} \leq (\frac{T}{T_c}-1) \leq 10^0$) for both the systems at all densities. The critical exponent $\gamma$ is obtained from 
the slope of the linear fit. For clarity, data at different densities are vertically shifted.
}}
\label{MCT_validy}
\end{figure}

  



\begin{figure}[h]
\centering
\includegraphics[width=0.45\textwidth]{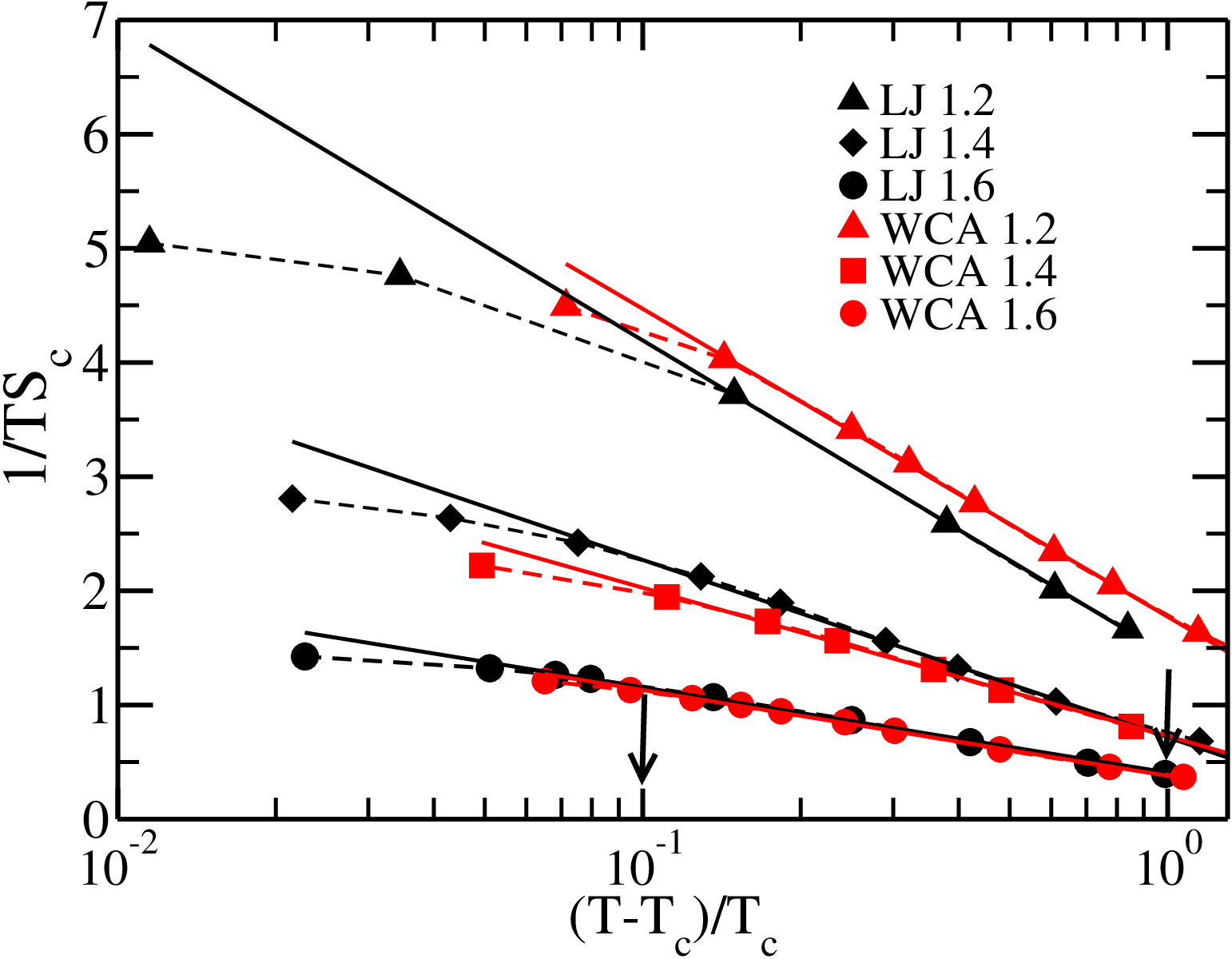}
\caption{ \it{ $\frac{1}{TS_c}$ plotted against $(\frac{T}{T_c}-1)$ and  as predicted by Eq.\ref{1_TSC} the plot is linear in the region $10^{-1} \leq (\frac{T}{T_c}-1) \leq 10^0$
validating our claim that MCT power law region overlaps with AG region.}  }
\label{TSC_validity}
\end{figure}

Although microscopic MCT shows a divergence of the relaxation time, $\tau$, at a much higher temperature \cite{Reichman} than the glass transition temperature, the power law behaviour of $\tau$ as predicted by MCT 
is found to be valid in a range of low temperatures. Similar to the earlier studies \cite{szamel-pre,tarjus_pre},
the power law behaviour of simulated $\tau$ we compute is well described by an algebraic divergence given by,
\begin{eqnarray}
 \tau\sim (T-T_c)^{-\gamma} \sim (\frac{T}{T_c}-1)^{-\gamma}
\label{t-tc}
\end{eqnarray}

For all the densities we study, as shown in Fig.\ref{MCT_validy}, in a certain region of temperature, ($10^{-1} \leq (\frac{T}{T_c}-1) \leq 10^0$), 
the relaxation time ,$\tau$, for both LJ and WCA systems 
follow the MCT power law behaviour. 
 On the other hand, the
Adam Gibbs relation is also valid for all the systems in this region ($0 \lesssim (\frac{T}{T_c}-1) \leq 10^0$) \cite{manu_under_prep}.
Thus we find that the temperature range where MCT like behaviour is predicted
 completely overlaps with the range where Adam-Gibbs relation is found to be valid. 
 As mentioned in the Introduction this overlap regime has earlier been reported for other systems \cite{sciortino-pre-2002}.
As in this temperature regime $\tau_\alpha$ can be described both by MCT power law behaviour and and by the AG relation  we can write,
\begin{eqnarray}
 \frac{A}{TS_c} \propto -\gamma \ln(\frac{T}{T_c}-1) 
\label{1_TSC}
\end{eqnarray}
In Fig.\ref{TSC_validity} we show that $\frac{1}{TS_c}$ is linear when plotted against $\ln (\frac{T}{T_c}-1)$ in the region $10^{-1} \leq (\frac{T}{T_c}-1) \leq 10^0$
validating the statement that MCT like divergence region overlaps with AG region.
\begin{figure}[h]
\centering
\includegraphics[width=0.45\textwidth]{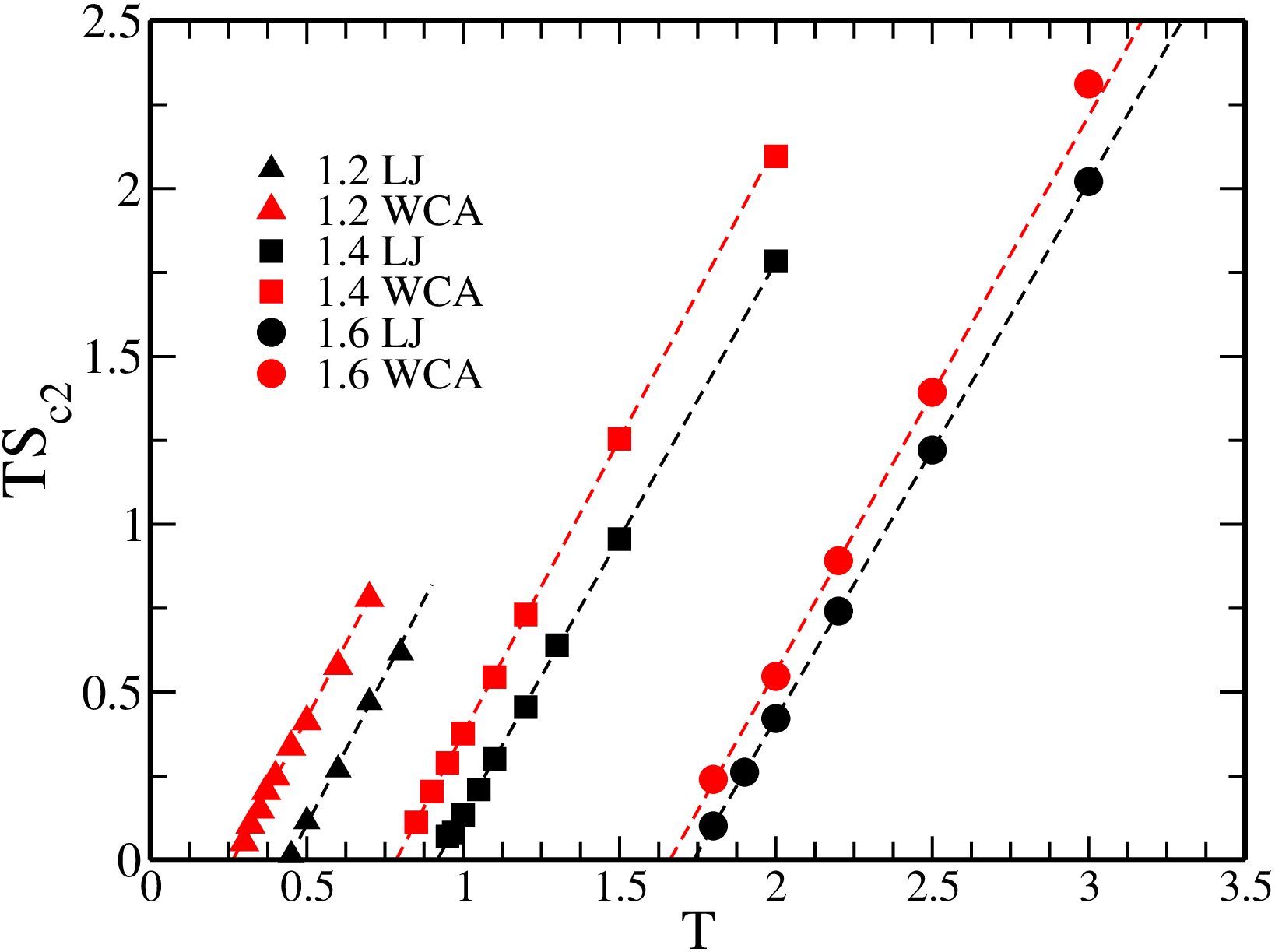}
\caption{
\it{ The temperature dependence of pair configurational entropy ($S_{c2}$) to determine  Kauzmann temperature $T_{K2}$. $T_{K2}$ values are given 
in Table \ref{tc_tk2}.  
}}
\label{TSC2_plot}
\end{figure}

 Since the configurational entropy has a finite value at the MCT transition temperature,
 $T_{c}$, the AG relation is not expected to predict a divergent relaxation time at this temperature. In order to investigate the origin of this avoided transition, we consider the separation of
 the configurational entropy into pair and many body parts as described earlier (sec-3.5)\cite{bssb}.
 We find that the temperature dependence of ($S_{c2}$) is given by (Fig.\ref{TSC2_plot}),

\begin{eqnarray}
 TS_{c2}=K_{T2}(\frac{T}{T_{K2}}-1) 
\label{pair_fragility}
\end{eqnarray}

\noindent where $K_{T2}$ is the pair thermodynamic fragility and $S_{c2}$ vanishes at the Kauzmann temperature $T_{K2}$ \cite{bssb}.
$T_{K2}$ is
obtained from the linear fit of $TS_{c2}$ vs $T$ plot at $S_{c2}=0$. As reported earlier we find that
for all the systems studied in this work the Kauzmann temperature for $S_{c2}$ is very close in value to the MCT transition temperature (Table \ref{tc_tk2}).

 \begin{table}[h]
 \centering
\caption{\it{$T_c$ \cite{tarjus_berthier_jcp} and $T_{K2}$ values are tabulated below. For all the systems studied here, 
 the Kauzmann temperature for $S_{c2}$ is quite 
similar to the MCT transition temperature.}}
\begin{tabular}{ l| r r |r r |r r}
 \hline
\hline
  & &$\rho=1.2$ && $\rho=1.4$ & $\rho=1.6$\\
\hline
& $T_{c}$& $T_{K2}$ & $T_{c}$& $T_{K2}$ & $T_{c}$ & $T_{K2}$
 \\
 \hline
  
LJ &0.435  &0.445 & 0.93 &0.929 & 1.76 &1.757\\
 WCA & 0.28& 0.268 & 0.81 &0.788 &1.69 &1.696\\
\hline
\hline
\end{tabular}
\label{tc_tk2}
\end{table}

Thus, although $S_{c}$ is finite at the estimated MCT $T_c$,  $S_{c2}$ 
vanishes at $T_{K2}$ which coincides with $T_c$. 


\section{Analytical Results}

Our study shows that the AG theory which is based on activation dynamics can completely describe the mode coupling theory (MCT) power law behavior
 in the region where the latter is found to be valid (Fig.\ref{TSC_validity}). However, the microscopic picture for Mode Coupling Theory
  (MCT) and the Adam Gibbs (AG) relation are different. Either from
  the heuristic arguments of Adam and Gibbs, or from the Random First
  Order Transition (RFOT) derivation, the AG relation is obtained from
  an activation picture of the dynamics, whereas the MCT does not
  correspond to activated dynamics. This leads to the question of the role of entropy in MCT which will be the focus of this section.

\subsection{Entropy and MCT}

In the $k \rightarrow 0$ limit, the memory function, $\mathcal{M}(k,t)$ in Eq.\ref{memory-fkt} can be rewritten as,
 
\begin{eqnarray}
\label{memory-fkt2}
 \mathcal{M}(k,t)=\frac{S(k)}{8\pi^2\rho k}\int_{0}^{\infty} dq \times q^2 (S(q)-1)^2\times \phi^2(q,t)
\end{eqnarray}
In the Schematic MCT  the $\phi(q,t)$ is usually decoupled from q, as in the memory function, $\mathcal{M}(k,t)$, the 
dominant contribution comes from the first peak of S(q) \cite{Bengtzelius,leutheusser}. 
Here we consider similar decoupling, however do not restrict ourself to first peak of S(q). Thus we write Eq.\ref{memory-fkt2} as
\begin{eqnarray}
\label{dcoupled-mfkt}
\mathcal{M}(k,t)=\frac{S(k)}{4\rho k (2\pi)^3}\left[\int_{0}^{\infty} d{\bf{q}} (S(q)-1)^2\right]\times \phi^2(k,t)
\end{eqnarray}
By writing S(q) in terms of g(r) we can rewrite Eq.\ref{dcoupled-mfkt} as 
\begin{equation}
\mathcal{M}(k,t)=\frac{S(k)}{2 k}\times 2\pi \rho \left[\int dr r^2(g(r)-1)^2 \right]\phi^2(k,t)
\label{fkt-s2-ap}
\end{equation}
Replacing $\mathcal{M}(k,t)$ from Eq.\ref{fkt-s2-ap} in Eq.\ref{fkt-eqn} and considering over damped limit by omiting the explicit `k' 
dependence of $\phi(t)$, Eq.\ref{fkt-eqn}
can be written in schematic form as 
\begin{equation}
 \label{sch-fkt}
\dot \phi(t) +\Omega^2 \phi(t) +\Omega^2 \lambda \int_0^t dt' \phi^2(t')\dot \phi(t-t') =0
\end{equation}
Where we can identify the coupling parameter $\lambda$ from Eq.\ref{fkt-s2-ap} as
\begin{equation}
\lambda=\frac{S(k)}{2k}\times 2\pi \rho \int dr r^2(g(r)-1)^2 =-\frac{S(k)}{2k}\frac{S_{2approx}}{k_B}
\label{lambda-fkt}
\end{equation}
Where we call $S_{2approx}$ as the approximate pair entropy. The choice of calling it entropy will become clear in the next analysis.

We note that the two body pair entropy is given by \cite{green_jcp},
\begin{equation}
\frac{S_{2}}{k_B}=-2\pi \rho  \int_0^{\infty}dr r^2 \{g(r) \ln g(r)- [g(r)-1]\}
\label{s2}
\end{equation}

\begin{figure}[h]
\centering
\includegraphics[width=0.45\textwidth]{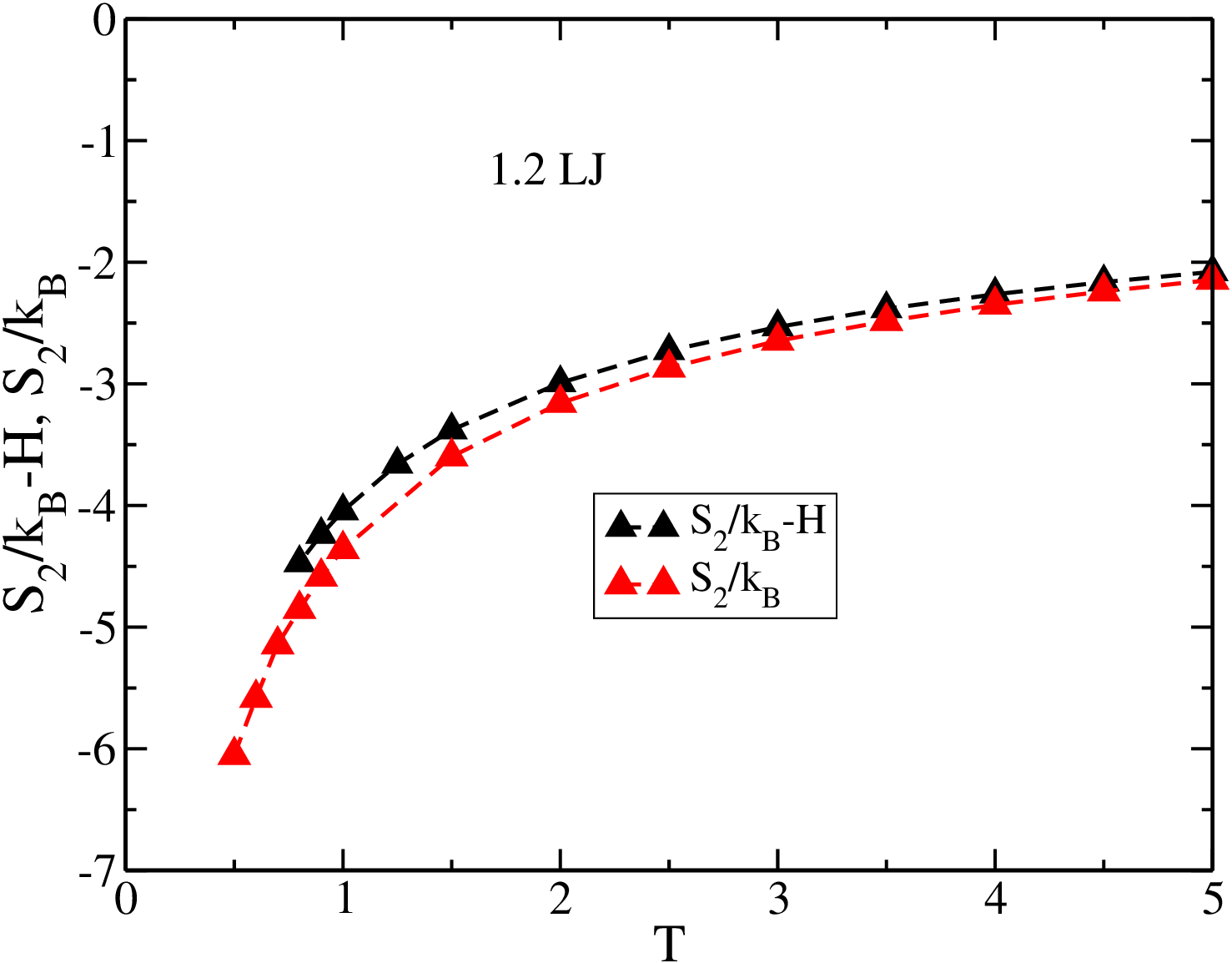}
\caption{ \it{The function ($\frac{S_2}{k_B}-H$) of Eq.\ref{s2-expanded} and $S_2/k_{B}$ are plotted as a function of temperature. The plot shows that the 
primary contribution
comes from the first term of the expansion. }}
\label{s2-s2-expand}
\end{figure}

Expanding the logarithmic term for $g(r) > 0$ we get
\begin{equation}
 \frac{S_{2}}{k_B}=-2\pi \rho  \int_0^{\infty} dr r^2 [g(r)-1]^2\frac{1}{(g(r)+1)}+H
\label{s2-expanded}
\end{equation}
where in `H' we put the higher order contributions. The Fig.\ref{s2-s2-expand} shows that the primary contribution comes from
the first term of Eq.\ref{s2-expanded}. 

\begin{figure}[h]
\centering
\includegraphics[width=0.45\textwidth]{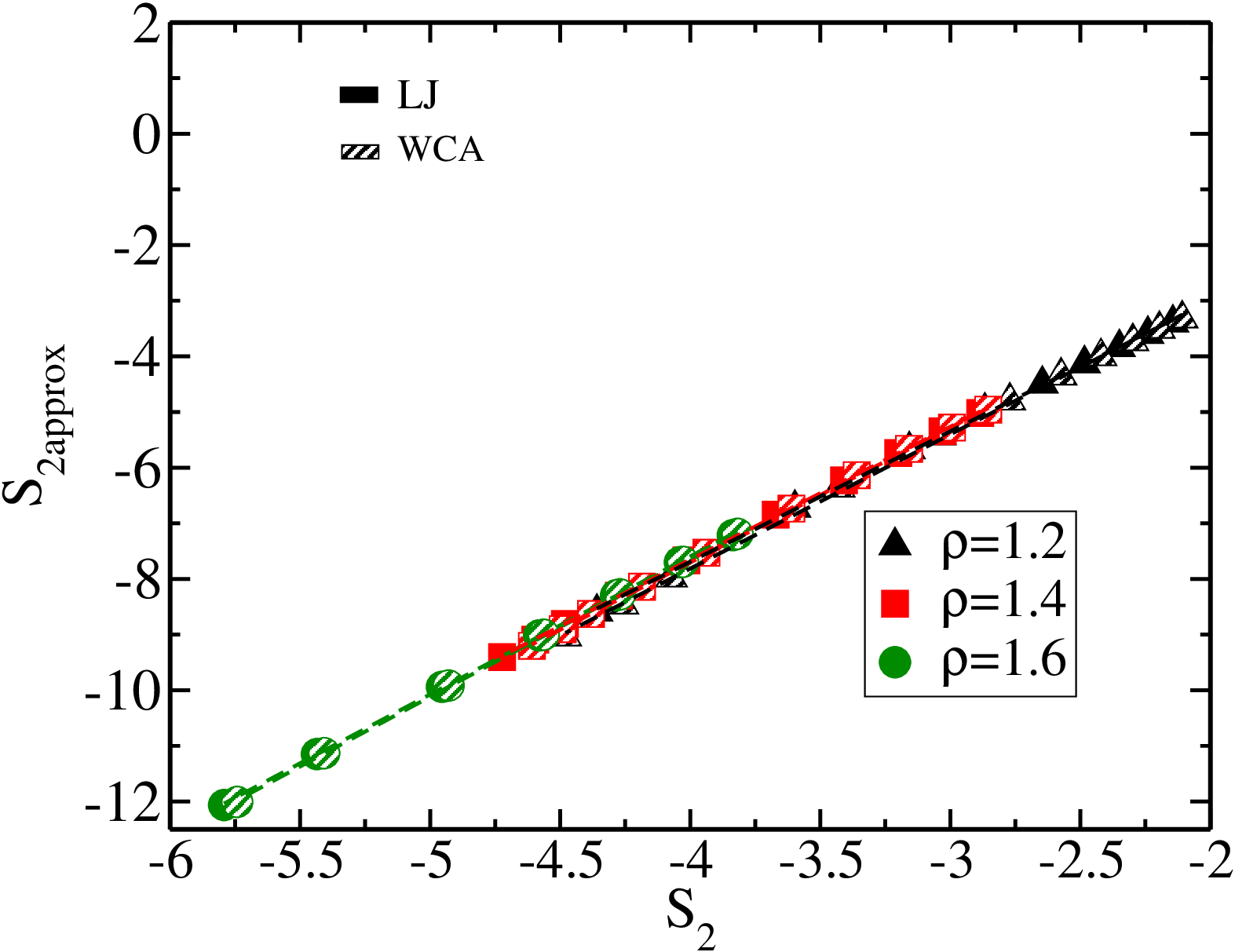}
\caption{ \it{$S_{2approx}$ is plotted against $S_{2}$ and it shows a linear behaviour with a slope $\approx 2.5$.}}
\label{fig-s2-s2app}
\end{figure}

In the above equation we note that $r^2[g(r)-1]^2$ varies strongly compared to $1/(g(r)+1)$. In the later if we consider $g(r)\approx 1$
we can write
\begin{equation}
 \frac{S_{2approx}}{k_B}=-2\pi \rho \int_0^{\infty} dr r^2 [g(r)-1]^2 \sim 2\frac{S_2}{k_B}-2H
\end{equation}
Our numerical analysis shows that for all the systems studied here $S_{2approx}$ vs $S_2$ is indeed linear (Fig.\ref{fig-s2-s2app})
with a slope $\approx 2.5$.
Thus the coupling constant $\lambda$ is related to the pair entropy,
\begin{equation}
\label{lambda-s2app}
 \lambda=-\frac{S(k)}{2k }\times \frac{S_{2approx}}{k_B}=-m_s\frac{S(k)}{2k}(S_2/k_B-H)
\end{equation}
where $m_s$  is the slope obtained from $S_{2approx}$ vs $S_2$ plot.

The MCT relaxation time from schematic model \cite{leutheusser} is given by
\begin{equation}
\label{lambda_gamma}
 \tau \sim (1-\lambda)^{-\gamma}
\end{equation} 
Note that the power law behaviour of relaxation time $\tau$ (as given by Eq.\ref{lambda_gamma}) changes to exponential dependence 
of $\tau$ under generalized MCT formalism\cite{jansen-reichman}, when the coupling parameter is considered to be the same for all
higher order terms and frequency $\Omega \sim 1$.
With these conditions $\tau$ can be written as
\begin{equation}
 \label{tau-lambda}
 \tau=\frac{1}{\Omega^2 \lambda}(exp(\lambda)-1)\sim \frac{\exp(\lambda)}{\lambda}\sim C'\exp(K' S_2)
\end{equation}
The second equality is written by replacing $\lambda$ from Eq.\ref{lambda-s2app}.
Where $C'$ and $K'$ are not a constants, rather have a temperature dependence.\\

Earlier study of diffusion \cite{alok} and our present microscopic derivation of the Rosenfeld relation for relaxation time $\tau$ shows that similar
to Rosenfeld prediction, the MCT also predicts it to be an universal scaling law for all transport coefficients.

\section{Numerical Results}

\subsection{Rosenfeld scaling and MCT}

\begin{figure}[h]
\centering
\subfigure{
 \includegraphics[width=0.45\textwidth]{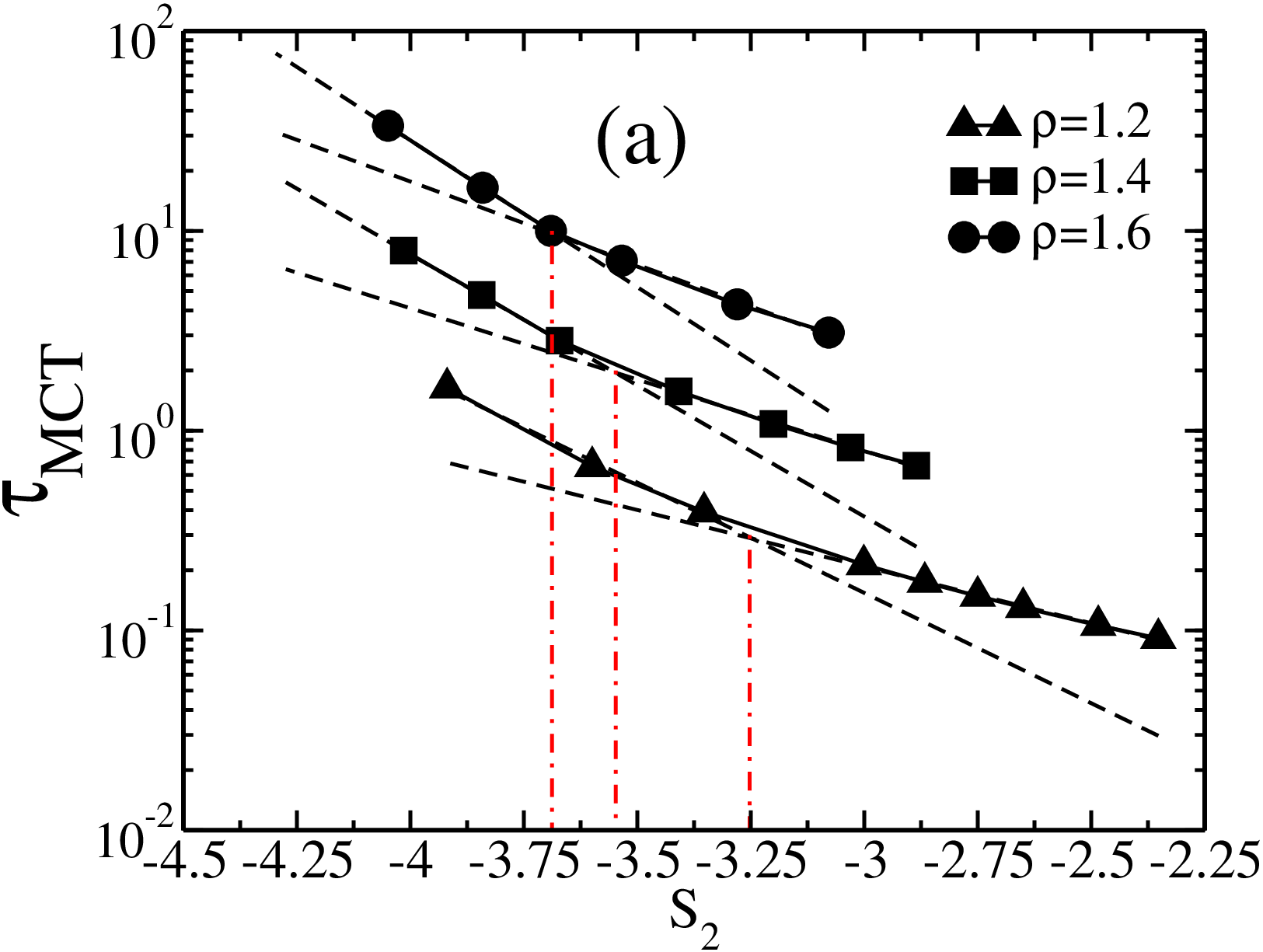}}
\subfigure{
 \includegraphics[width=0.45\textwidth]{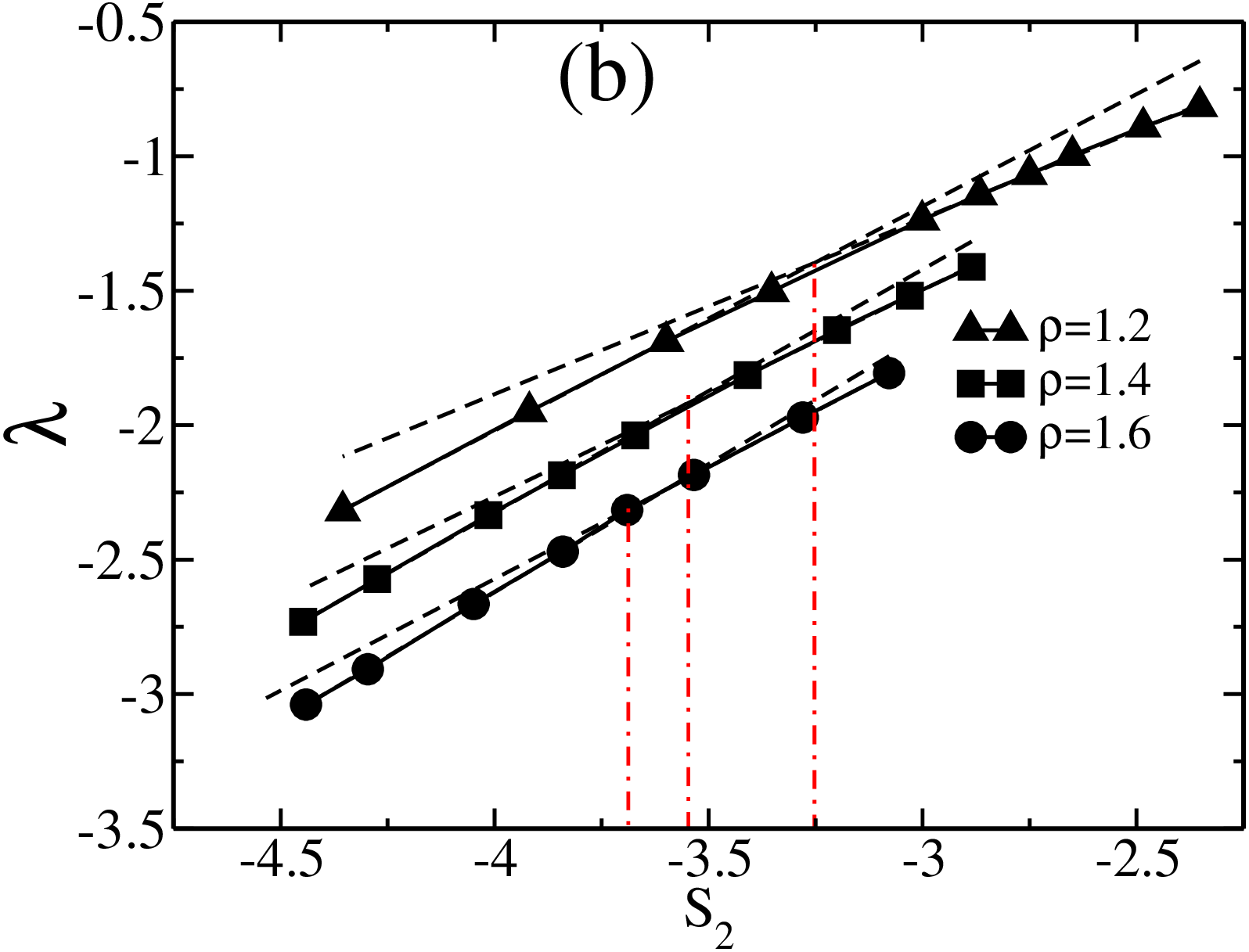}}
\caption{\it{(a) The relaxation time obtained from microscopic MCT, $\tau_{MCT}$ is plotted against $S_2$. The dashed lines illustrate the two different Rosenfeld regimes.
(b) Plot of  $\lambda$ vs. $S_2$. This also shows two different linear regimes.
For clarity $\tau$ and $\lambda$ are shifted by 1.2, 2.8 and -0.2, -0.48 for the systems of $\rho=1.4$ and $\rho=1.6$ respectively.
The break in the slope for both the plots are illustrated by vertical dash-dot lines. We show that for a fixed density the break for both $\tau_{MCT}$ and $\lambda$ are at the same $S_2$ value.}}
\label{tau-sexc}
\end{figure}

In this section we analyze the MCT results in the light of Rosenfeld relation. We find that the relaxation time as obtained
from microscopic MCT, $\tau_{MCT}$ when plotted against $\lambda$ does not follow the power law ($(1-\lambda)^{-\gamma}$) or
$exp(\lambda)$ dependence in the whole temperature region. Usually it is found \cite{Ruchi_charu_2006,shankar_das} that both
$\tau_{MCT}$ and $\tau$ (relaxation time obtained from simulation) when plotted against $S_2$ 
 does not show a single straight line. In Fig.\ref{tau-sexc} we plot the $\tau_{MCT}$ calculated from Eq.9-12 against $S_2$ which shows two linear regimes.
 The origin of this break or the temperature dependence of
the Rosenfeld parameter `$K'$' is not known.

Our analysis of Eq.\ref{lambda-s2app} shows that the Rosenfeld parameters are related to the static structure factor $S(k)$. Thus the temperature dependence of $S(k)$ leads to 
the temperature dependence of Rosenfeld parameter `$K'$'. However since $S(k)$ changes continuously with temperature, it should lead to a similar temperature dependence of $K'$.
That a continuously changing `$K'$' is not needed to describe the observed behaviour 
but two distinct values  suffice can be seen when we plot
 $\lambda$ against $S_2$ (Fig.\ref{tau-sexc}-b), where we see that 
there is a break in the slope and it happens at the same $S_2$ value where $\tau$ against $S_2$ shows a break in slope.

\begin{figure}[h]
\centering
\subfigure{
\includegraphics[width=0.4\textwidth]{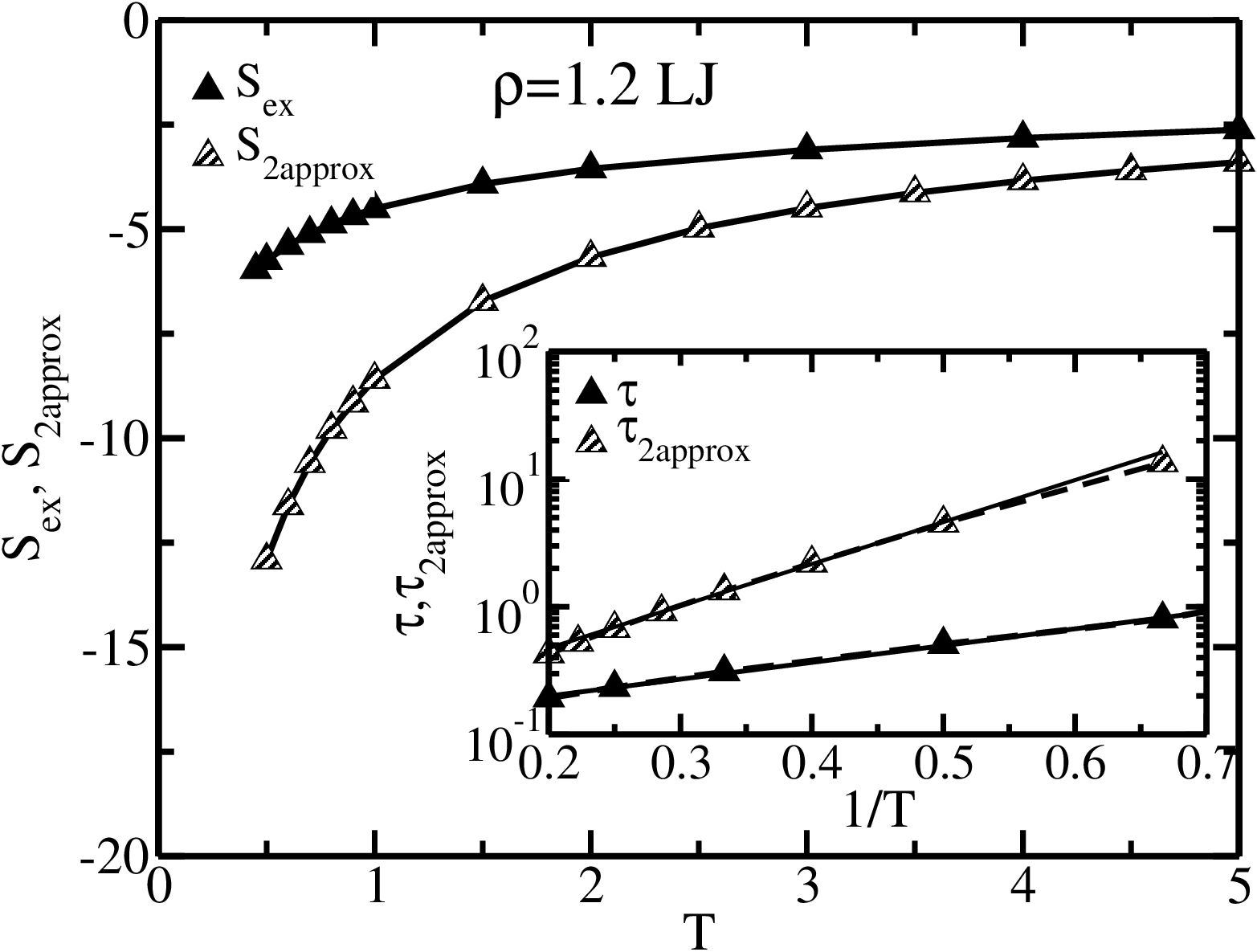}}
\subfigure{
\includegraphics[width=0.4\textwidth]{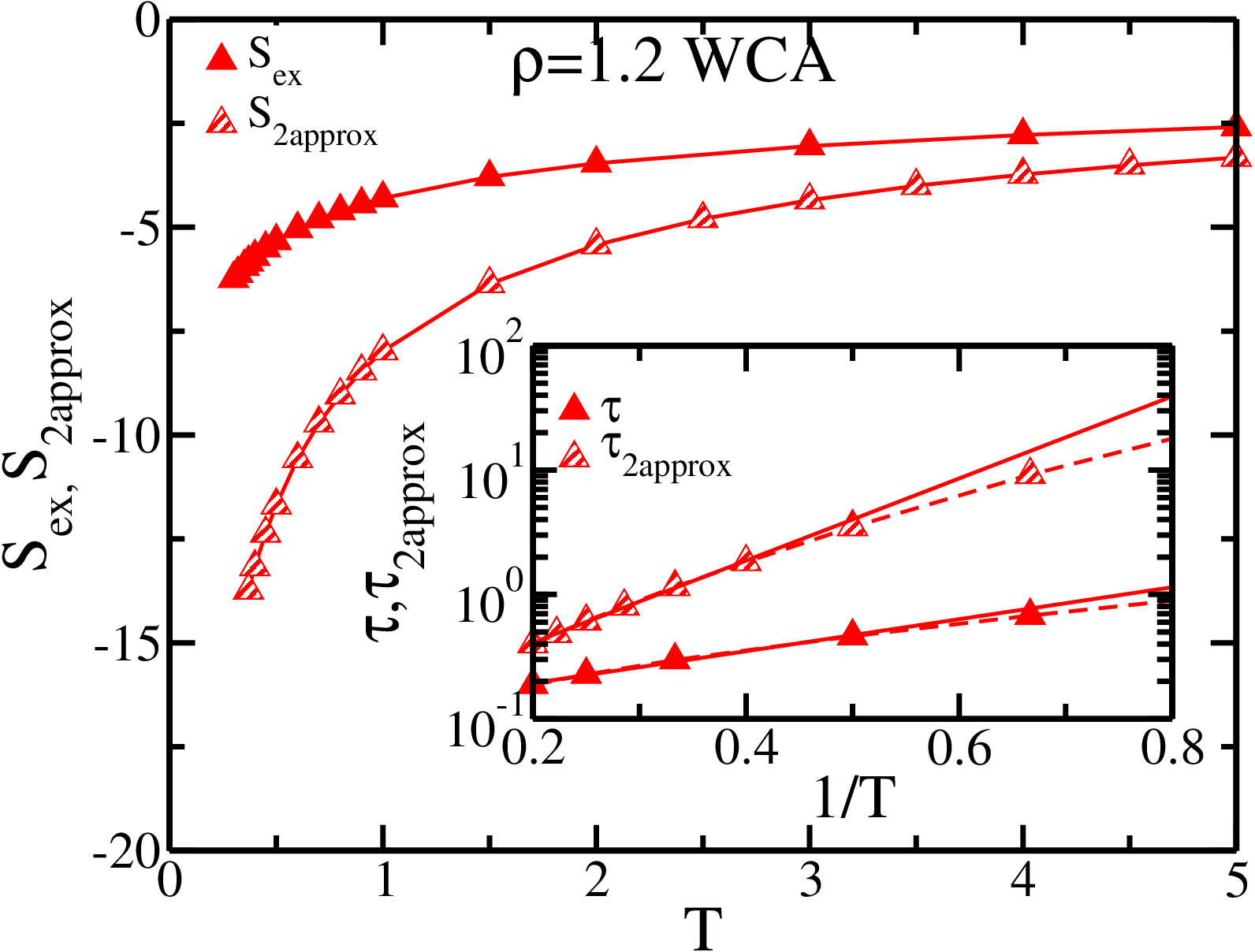}}
\subfigure{
\includegraphics[width=0.4\textwidth]{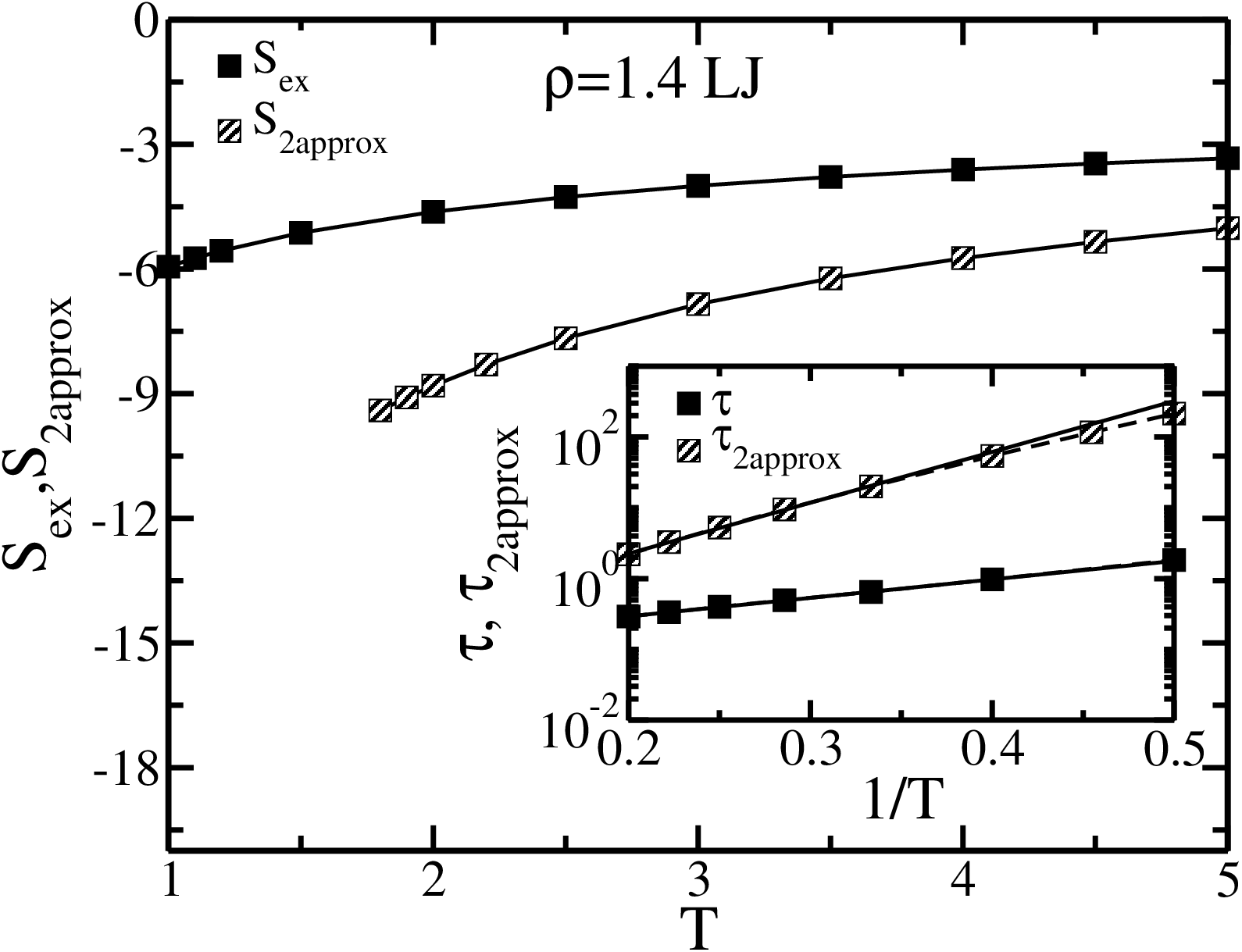}}
\subfigure{
\includegraphics[width=0.4\textwidth]{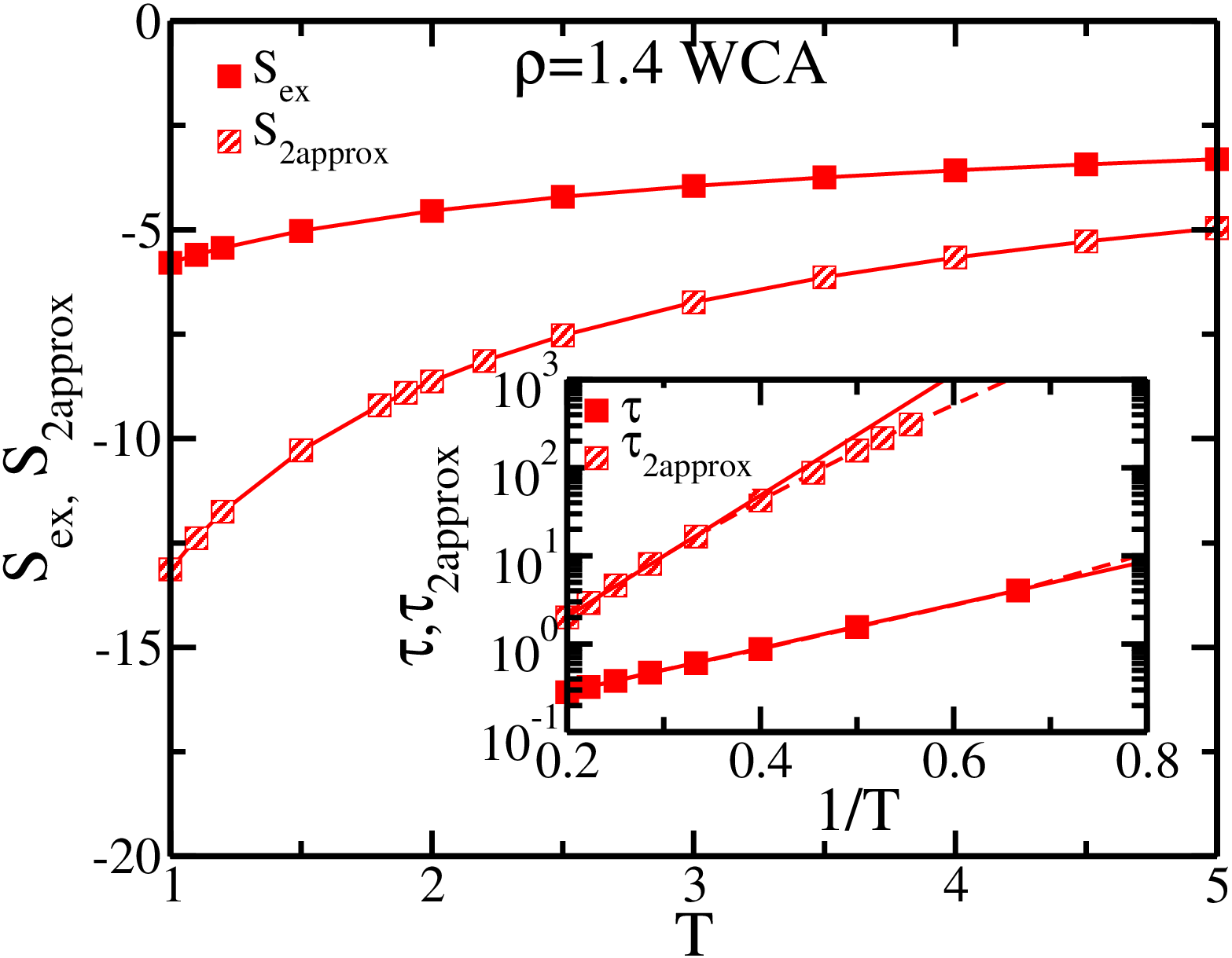}}
\subfigure{
\includegraphics[width=0.4\textwidth]{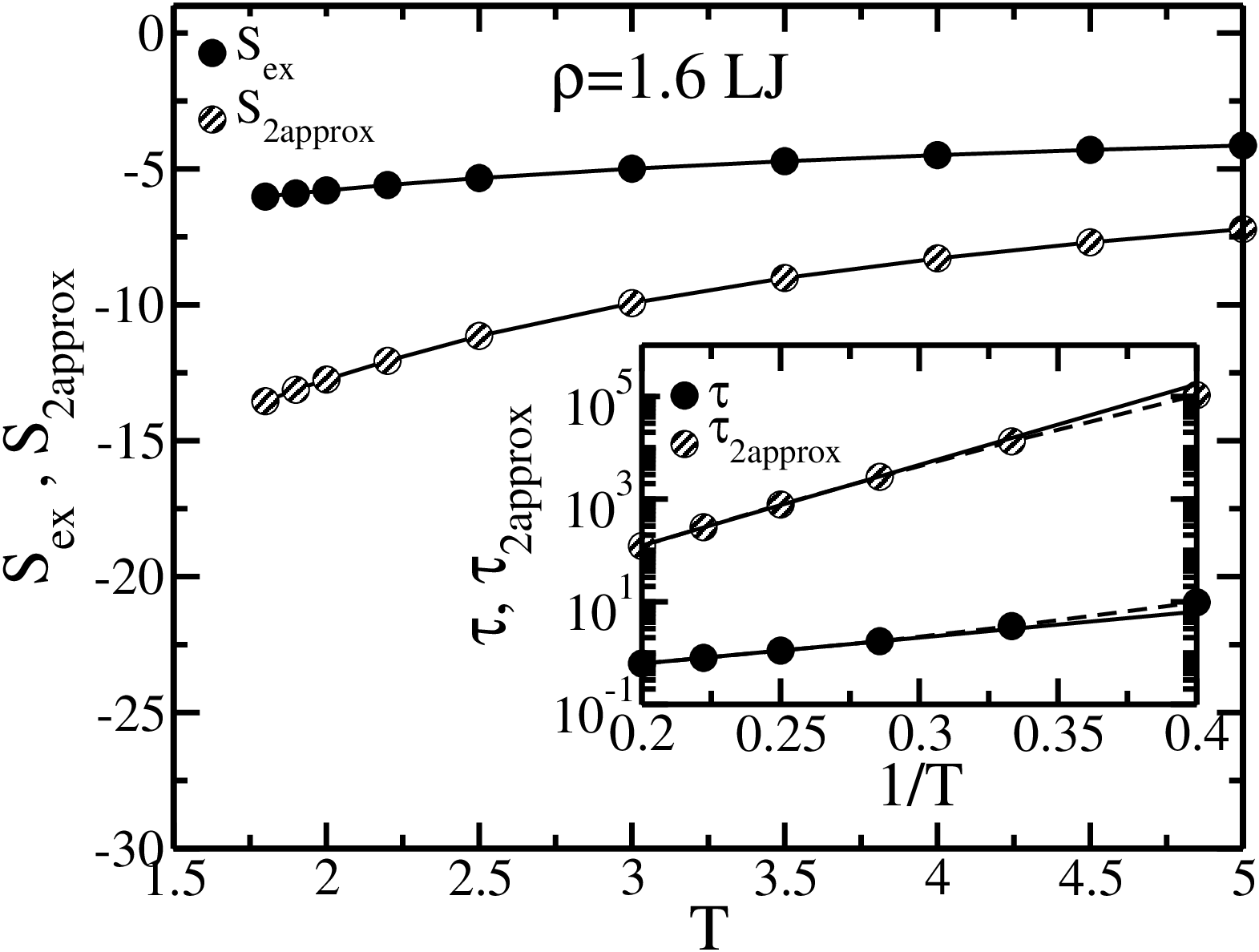}}
\subfigure{
\includegraphics[width=0.4\textwidth]{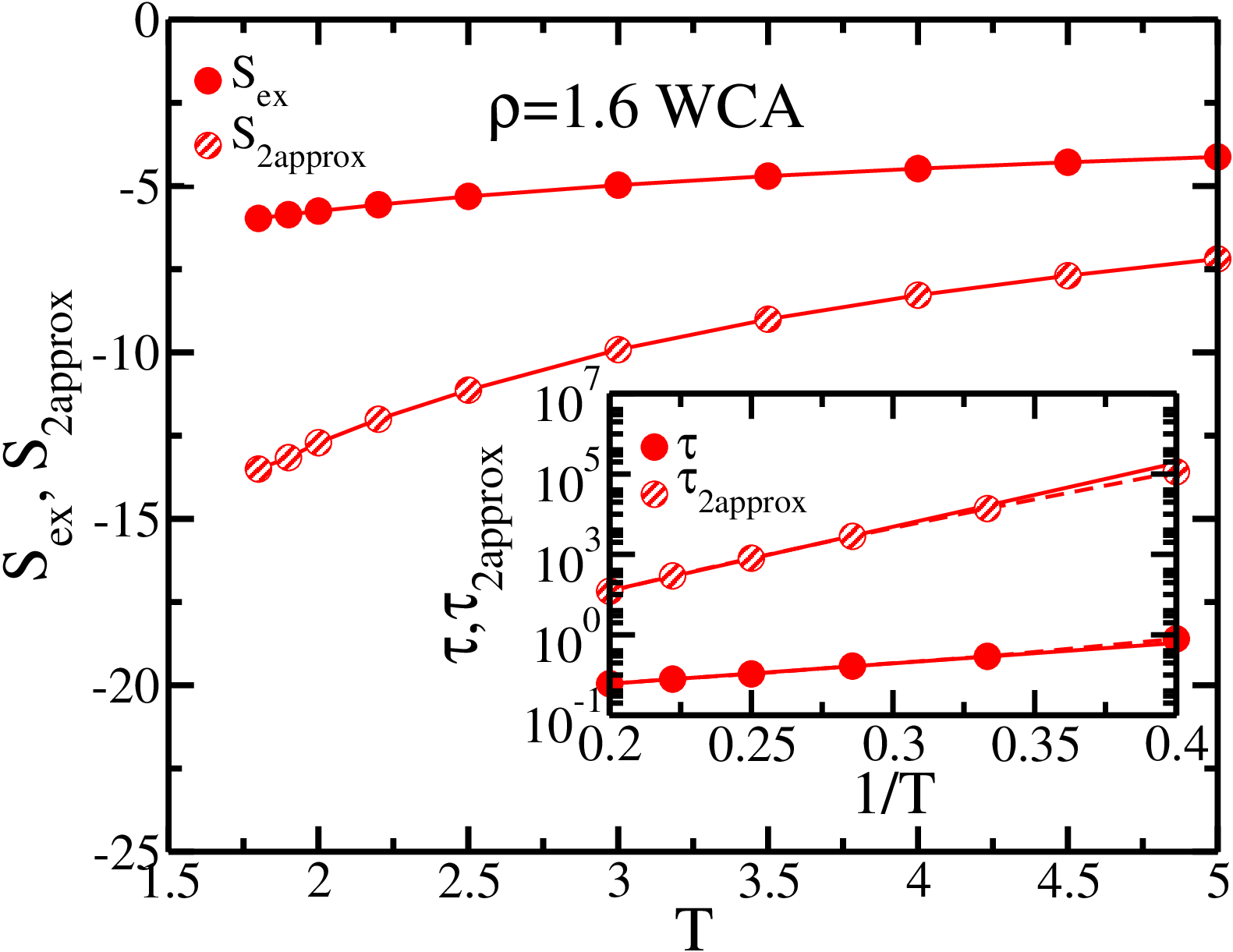}}
\caption{\it{$S_{ex}$ and $S_{2approx}$ are plotted as a function of temperature for LJ and WCA systems at densities 1.2, 1.4 and 1.6. For all the systems
$S_{2approx}$ has stronger temperature dependence and has smaller value than $S_{ex}$. In inset we plot $\tau_{2apporx}$ as obtained from Eq.\ref{tau-approx}. It 
shows that $\tau_{2approx}$ has higher value and a larger slope 
leading to higher activation energy as compared to $\tau$. Activation energies are tabulated in Table \ref{e0table}}}
\label{fig-sexc-s2-tau}
\end{figure}

Next we show that the value of  $S_{2approx}$ and its temperature dependence as compared to $S_{ex}$ can explain i) the larger 
values of $\tau_{MCT}$ as compared to $\tau$ \cite{szamel-pre} ii) the higher values of activation energy as predicted by MCT 
\cite{tarjus_pre}. When $E_0^{sim}$
and $E_0^{MCT}$ are obtained by fitting $\tau$ and $\tau_{MCT}$ to Arrhenius expression (Eq.\ref{arhenius-eq})  we find values shown in Table 2, and in Fig.7. 
\begin{eqnarray}
 \tau \sim \tau_{0}\exp \frac{E_{0}}{T}
\label{arhenius-eq}
\end{eqnarray}
 
\begin{table}[h]
 \centering
\caption{\it{$E_0$ are tabulated for different systems. We show that the $E_0$ values are higher for MCT as well as for approximate
calculation. As we can not calculate overlap function from MCT, for comparison of $E_0$ values we use simulated $F_s(k,t)$}}
\begin{tabular}{ l |r r |r r |r r}
\hline
\hline
  & &$\rho=1.2$ && $\rho=1.4$ & $\rho=1.6$\\
\hline
& LJ& WCA & LJ&WCA & LJ& WCA\\
 \hline 
$E_0^{sim}$ &2.509 & 1.901 & 5.997 & 5.694 & 12.499 &11.749\\
$E_0^{MCT}$ & 5.002& 3.993 & 11.565 & 10.775 &21.748 &21.082\\
$E_0^{approx}$ & 6.224& 5.705 & 16.535 & 15.831 &37.159 &36.564\\
\hline
\hline
\end{tabular}

\label{e0table}
\end{table}

Fig.\ref{fig-sexc-s2-tau} shows that at all densities for both the systems $S_{2approx}$ is smaller than $S_{ex}$ and has a much 
stronger temperature dependence. Using Rosenfeld Expression we can write
\begin{equation}
\tau(T)=C \exp(-K S_{ex})
\end{equation}
Now if we replace $S_{ex}$ by $S_{2approx}$, keeping C and $K$ same, we get 
\begin{equation}
 \tau_{2approx}=C\exp(-K S_{2approx})
\label{tau-approx}
\end{equation}
The C and $K$ are obtained from linear fits of logarithmic of simulated relaxation time against excess entropy.
Since $S_{2approx} << S_{ex} $, the study shows that $\tau_{2approx} >>\tau $. 
Similar to that predicted by microscopic MCT (Eq.\ref{fkt-eqn}, \ref{fskt-eqn}), the $E_0$ values for $\tau_{2approx}$ are higher, 
which are given in Table \ref{e0table}.

Although the results obtained from $\tau_{2approx}$ shows the correct trend, it can not match the parameters as obtained from 
$\tau_{MCT}$. We note that the $\tau_{2approx}$ is a prediction obtained from schematic MCT, which is known to 
overestimate the coupling constant $\lambda$.  However this  analysis not only explains the behaviour of MCT at high temperature, it
also throws some light in the origin of its breakdown at low temperature.  Usually the breakdown of MCT at low temperature has been attributed to the neglect 
of higher order correlation functions \cite{szamel-gmct,jansen-reichman}. This present analysis predicts that the stronger temperature dependence of the vertex
might be partially responsible for the breakdown of MCT even at low temperature.


\subsection{The Adam Gibbs Relation and MCT}

We have shown that the relaxation time ,$\tau$, over a temperature regime ($10^{-1} \leq (\frac{T}{T_c}-1) \leq 10^0$) follows both the AG relation and MCT power
 law behaviour. We also find the avoided
divergence obeserved in the configurational entropy plot (Fig.\ref{TSC_validity}) arises from the vanishing of the pair configurational entropy ($S_{c2}$). For all 
the systems studied here, we find $T_{K2} \simeq T_c$ (Table \ref{tc_tk2}), thus we can rewrite Eq.\ref{pair_fragility} as,
\begin{eqnarray}
TS_{c2}=K_{T2}(\frac{T}{T_{K2}}-1)\simeq K_{T2}(\frac{T}{T_{c}}-1) 
\label{T_sc_Tc}
\end{eqnarray}

We note that although $T_{K2} \simeq T_c$ and the MCT framework which predicts the power law behaviour is developed at the two body level, the AG relation 
with $S_{c2}$ alone cannot predict the MCT power law behaviour. and the RMPE ,$\Delta S$, plays an important role in predicting it. We also show that indeed
 there is a  relation between MCT critical exponent $\gamma$, Adam Gibbs coefficient A, the pair thermodynamic fragility $K_{T2}$.

As shown earlier in Eq.\ref{sc2}, configurational entropy can be written in terms of pair configurational entropy and RMPE. Thus we can write,

\begin{eqnarray}
\frac{A}{TS_c}=\frac{A}{TS_{c2}+T \Delta S}=  \frac{A}{K_{T2}}\frac{1}{[(\frac{T}{T_c}-1)+\frac{T \Delta S}{K_{T2}}]}
\label{A_SC2}
\end{eqnarray}
\noindent where we have used Eq.\ref{sc2} and Eq.\ref{T_sc_Tc} to write the first and second equality respectively.

\begin{figure}[h]
\centering
\subfigure{
\includegraphics[width=0.45\textwidth]{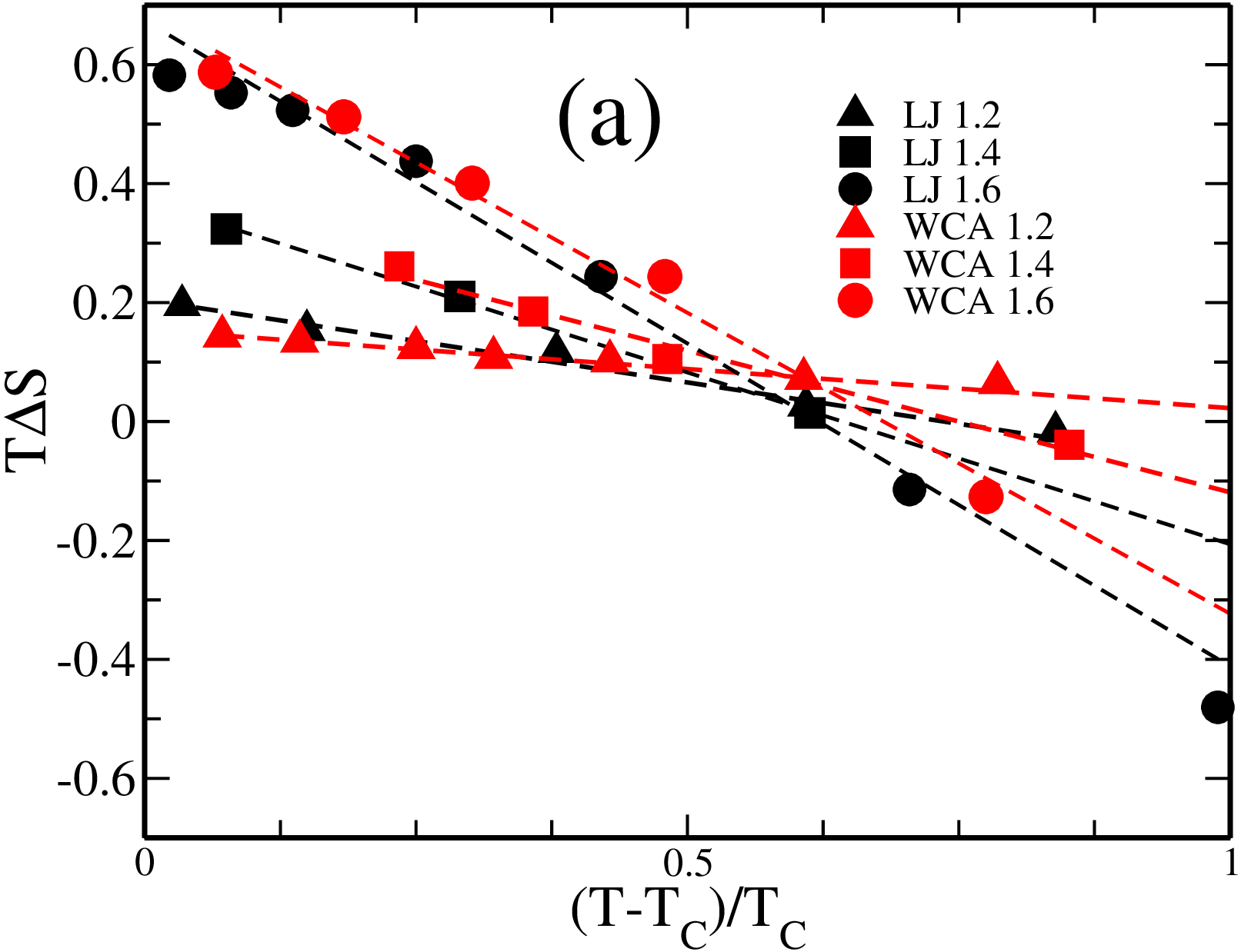}}
\subfigure{
\includegraphics[width=0.45\textwidth]{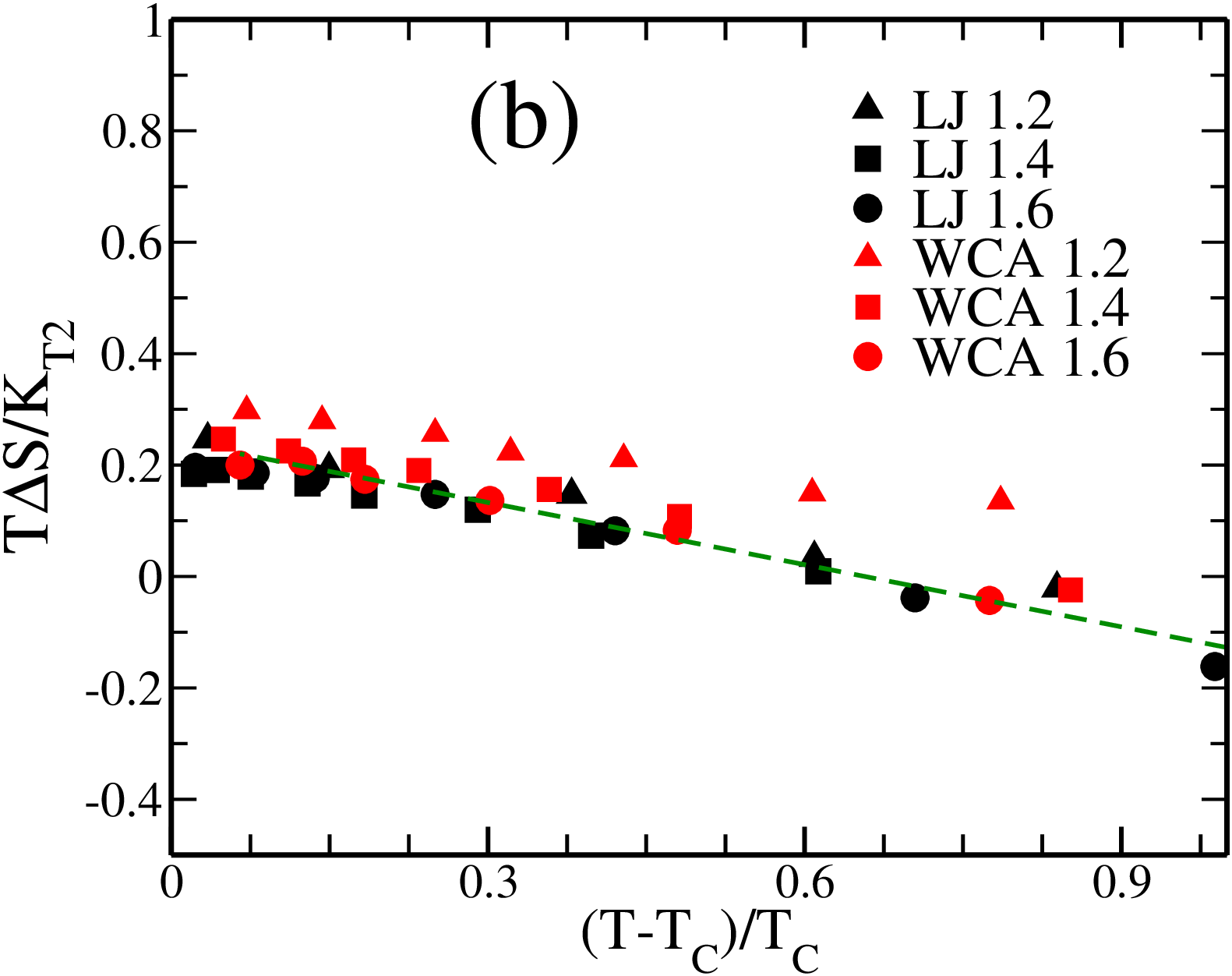}}
\subfigure{
\includegraphics[width=0.46\textwidth]{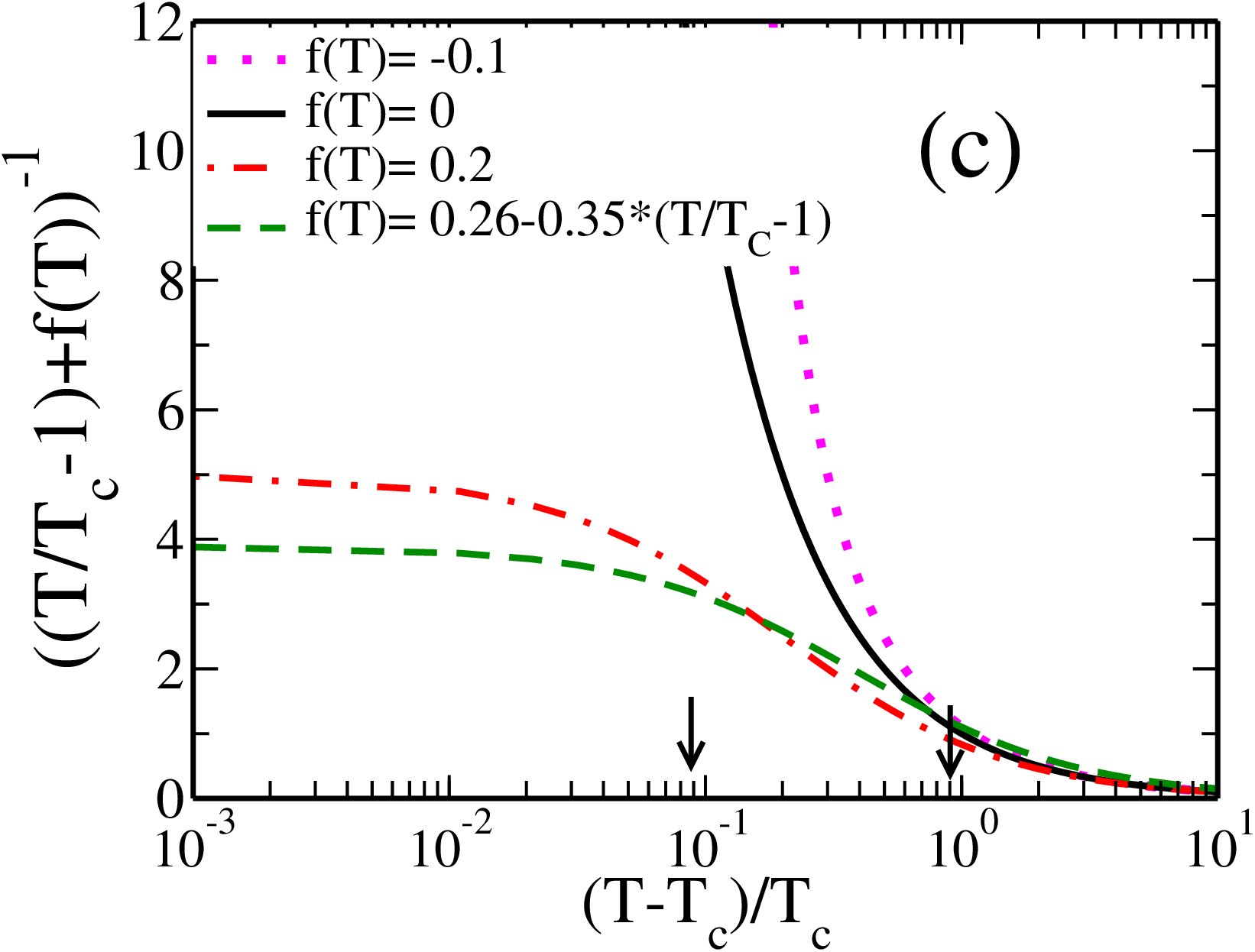}}
\subfigure{
\includegraphics[width=0.45\textwidth]{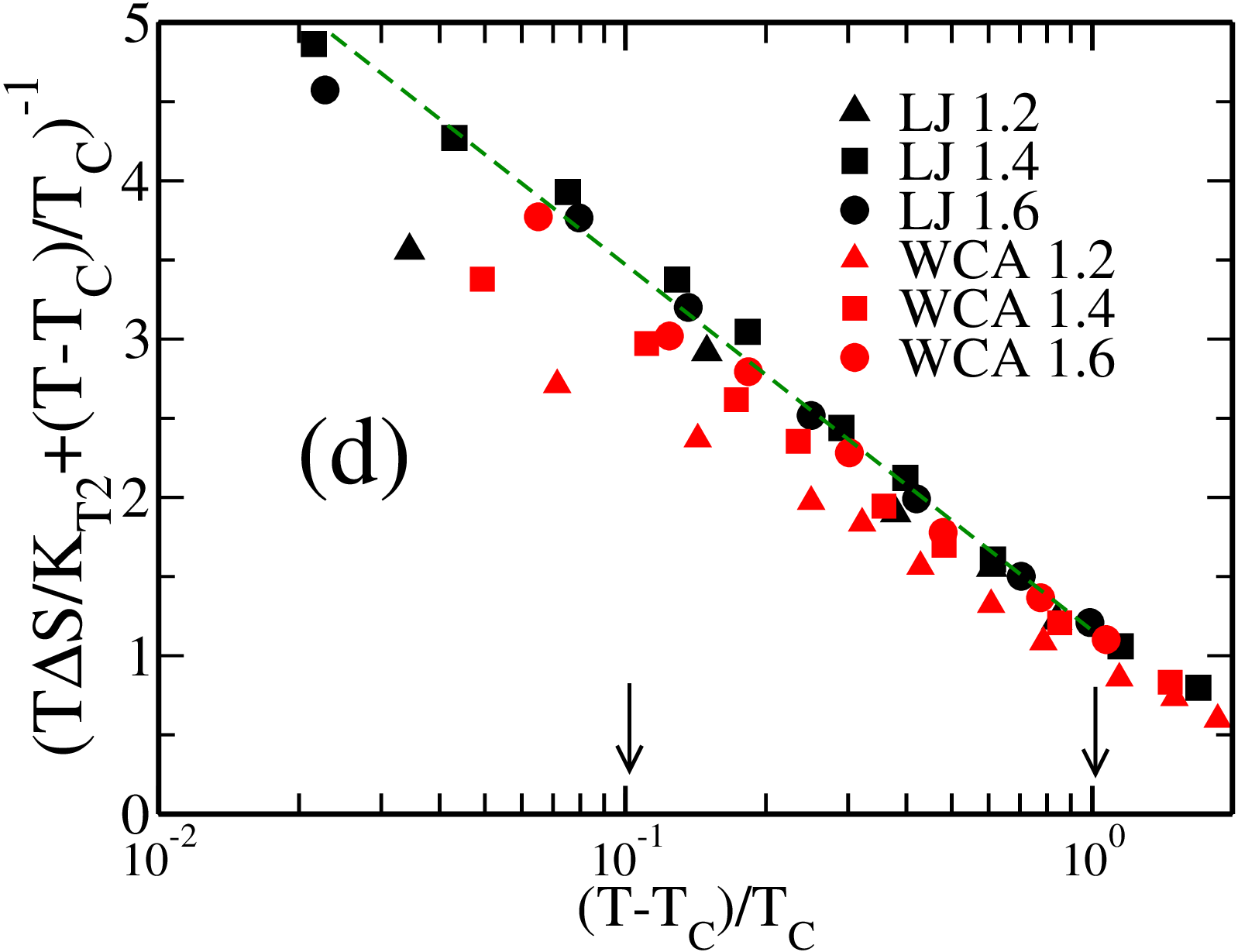}}
\caption{\it{ (a) $T\Delta S$ is plotted as a function of $(\frac{T}{T_c}-1)$ and it shows a strong  system dependence. (b)
  $\frac{T \Delta S}{K_{T2}}$ vs $(\frac{T}{T_c}-1)$  showing a master plot for all the systems except for
WCA system at $\rho=1.2$. Dotted line is guide to eye.
(c) $((\frac{T}{T_c}-1)+f(T))^{-1}$ plotted against $(\frac{T}{T_c}-1)$ by varying $f(T)$. 
Only for non negligible positive values of $f(T)$, linearity is found in the regime 0.1 to 1.0 of $ (\frac{T}{T_{c}}-1)$.
To obtain a correct estimation of the MCT 
power law exponent $\gamma$ (slope of the plot), f(T) needs to be temperature dependent (green dashed line).
(d) $[\frac{1}{(\frac{T}{T_c}-1)+\frac{T \Delta S}{K_{T2}}}]$ vs $(\frac{T}{T_c}-1)$ shows a master plot for all the systems except for
WCA system at $\rho=1.2$. `m' is the slope of the linear plot which is tabulated in Table \ref{slope}. 
}}
\label{delta_S}
\end{figure}

We find that although $T\Delta S$ is system dependent (Fig.\ref{delta_S}a), except for WCA system at $\rho = 1.2$ the function $\frac{T \Delta S}{K_{T2}}$ shows a master plot when plotted 
against $(\frac{T}{T_c}-1)$ (Fig.\ref{delta_S}b). 
 Note that although the value of  $\frac{T \Delta S}{K_{T2}}$ is small, it is not negligible.

The master plot of $\frac{T \Delta S}{K_{T2}}$ can be fitted to a straight line, $\frac{T \Delta S}{K_{T2}}=0.26-0.35(\frac{T}{T_c}-1)$.
Next we show that a function $\frac{1}{(\frac{T}{T_c}-1)+f(T)}$ when plotted against $\ln (\frac{T}{T_c}-1)$ shows linearity in the whole regime of
 $(10^{-1} \leq (\frac{T}{T_c}-1) \leq 10^0)$ only when $f(T)$ is non-negligible positive quantity (Fig.\ref{delta_S}c). 
 Note that in Fig.\ref{delta_S}c  when $f(T)=0$ (which implies $\Delta S=0$ in Eq.\ref{A_SC2}) the function diverges strongly.
This shows that the AG relation at two body level cannot predict the MCT power law behaviour. 

 The analysis further shows that to obtain a correct estimation of the MCT 
power law exponent $\gamma$ (slope of the plot), $f(T)$ needs to obey the following temperature dependence, $f(T)=\frac{T\Delta S}{K_{T2}}=0.26-0.35(\frac{T}{T_c}-1)$. 
The two functions $\frac{T\Delta S}{K_{T2}}$ and $(\frac{T}{T_c}-1)$  show opposite trends, the former increases whereas the later decreases with temperature.
 Therefore a crossover between
these two functions is observed in this regime and around MCT transition temperature, $\frac{T \Delta S}{K_{T2}}\gg (\frac{T}{T_c}-1)$ and 
configurational entropy and the relaxation time are determined primarily by many body contributions.

 From Fig.\ref{delta_S}d we find in the temperature regime ($10^{-1} \leq (\frac{T}{T_c}-1) \leq 10^0$) Eq.\ref{A_SC2} can be
re-written as,
\begin{eqnarray}
\frac{A}{TS_c}=\frac{A}{K_{T2}}\frac{1}{[(\frac{T}{T_c}-1)+\frac{T \Delta S}{K_{T2}}]} \sim -\frac{mA}{K_{T2}}\ln (\frac{T}{T_c}-1)
\label{A_SC2_new}
\end{eqnarray}
where `m' is the slope obtained from Fig.\ref{delta_S}(d) and given in Table \ref{slope}.
Since $\tau$ is found to follow AG relation we can write,
\begin{eqnarray}
 \tau \sim \exp(\frac{A}{TS_c}) \sim (\frac{T}{T_c}-1)^{\frac{mA}{K_{T2}}}
\label{tau_gamma_kt2}
\end{eqnarray}

\begin{table}
 \centering
\caption{ \it{ The slope of the linear plot of $[\frac{1}{(\frac{T}{T_c}-1)+\frac{T \Delta S}{K_{T2}}}]$ vs $(\frac{T}{T_c}-1)$ in the region $(10^{-1} \leq (\frac{T}{T_c}-1) \leq 10^0)$ 
(Fig.\ref{delta_S}d).}}
\begin{tabular}{c|c|c}
    \hline
    \hline
   $\rho$&$m(LJ)$ & $m(WCA)$  \\ \hline
    1.2&0.987& 0.695 \\ 
    1.4&1.029 & 0.888 \\ 
   1.6 & 1.004&1.000\\ \hline
 \hline

\end{tabular}
\label{slope}
\end{table}
\begin{table}[h]
\centering
\caption{\it{  $mA/K_{T2}$ and $\gamma$ for LJ and WCA system. As predicted by Eq.\ref{gamma_A_KT2_eq} $mA/K_{T2}$ value is similar to $\gamma$ value obtained
 from free fitting (Fig.\ref{MCT_validy} ) for most of 
the systems.}}
\begin{tabular}{ l |r r |r r |r r}
 \hline
\hline
  & &$\rho=1.2$ && $\rho=1.4$ & $\rho=1.6$\\
\hline
& $mA /K_{T2}$& $\gamma$ & $ mA/K_{T2}$& $\gamma$ &  $ mA/K_{T2}$& $\gamma$
 \\
 \hline
  LJ &2.322 &2.229 & 2.474 &2.385 & 2.352 &2.299\\
 WCA & 2.932& 2.243 & 2.852&2.289 &2.579 &2.304\\
\hline
\hline
\end{tabular}
\label{gamma_A_KT2}
\end{table}

\begin{table}[h]
\centering
\caption{\it{ The Adam Gibbs coefficient `A', as obtained
from the linear fit of $\tau$ vs $1/TS_c$ plot, and pair thermodynamic fragility $K_{T2}$, as obtained from the linear fit of $TS_{c2}$ vs $T/T_{k2}$ plot
 for both the systems at different densities is tabulated below. The data shows that both are strongly dependent on density.}}
\begin{tabular}{c|c|c|c|c}
    \hline
\hline
    $\rho$&$A(LJ)$ & $A(WCA)$ & $K_{T2}(LJ)$ & $K_{T2}(WCA)$  \\ \hline
    1.2&1.87 & 1.89 & 0.795&0.483 \\ \hline
    1.4&3.57 & 4.37 & 1.555 & 1.358 \\ \hline
    1.6 & 6.96&7.57& 2.971&  2.936\\ \hline
\hline
\end{tabular}
\label{A_value}
\end{table}
Comparing Eq.\ref{t-tc} and Eq.\ref{tau_gamma_kt2}  we can write, 
\begin{eqnarray}
 \frac{mA}{K_{T2}} \sim \gamma
\label{gamma_A_KT2_eq}
\end{eqnarray}
\noindent where $m\simeq 1$ for all the systems except for the WCA system at $\rho=1.2$.
 Thus we show that the MCT scaling parameter, $\gamma$ is related to the AG parameter,
 A and the pair thermodynamic fragility of $S_{c2}$, $K_{T2}$.
We have tabulated the $\gamma$ values in Table \ref{gamma_A_KT2}, which shows the above relation holds. The deviation of slope value (`m') from unity for WCA system
at $\rho=1.2$ may have some connection to its breakdown of density-temperature scaling which needs to be investigated in future.

  The MCT critical exponent ($\gamma$) is known to be density-temperature independent \cite {Gotze}. Interestingly we also find that although both AG coefficient (A) and pair thermodynamic fragility 
($K_{T2}$) are strongly dependent on density and temperature (Table \ref{A_value}), but their ratio, which is related to $\gamma$ (Eq.\ref{gamma_A_KT2_eq}), is 
density-temperature independent (Table \ref{gamma_A_KT2}).

\section{Conclusion}
In this work we show that in a certain region $(10^{-1} \leq (\frac{T}{T_c}-1) \leq 10^0)$ the relaxation time follows both the AG relation and MCT power law behaviour.
We also  find that the MCT divergence temperatures coincide with the temperature where
pair  configurational entropy goes to zero for all the systems studied here.
AG relation is based on activated dynamics, whereas MCT is mean field theory which at the two body level does not address any activated dynamics.
 Also the microscopic MCT does not have any apparent 
connection to entropy. Thus to understand the above mentioned observations   
we explore the connection between mode coupling theory  and entropy and discuss different predictions of MCT in the light of entropy.

In this article we show that the MCT vertex for the structural relaxation time under certain approximations can be related to the pair excess entropy.  Higher order MCT calculations in 
the schematic MCT framework can relate the relaxation time to the exponential of this vertex. Thus the MCT can provide a microscopic derivation of the phenomenological
 Rosenfeld theory. Our analysis shows that the Rosenfeld parameter is related to the
 static structure factor $S(k)$.  The temperature dependence of $S(k)$ leads to 
the temperature dependence of Rosenfeld parameter `K', thus explaining the earlier observation of the non-uniqueness of the Rosenfeld exponent \cite{shankar_das,manish_charu}.
The analysis of the vertex reveals that quantity which contributes to the vertex, $S_{2apporx}$ has a much lower value and stronger temperature dependence as compared 
to the excess entropy , $S_{ex}$.  If we assume the Rosenfeld scaling to be valid and replace $S_{ex}$ by $S_{2approx}$, the predicted relaxation time shows similar
 characteristics as the MCT relaxation time.  Thus the study reveals that the larger value of  $\tau_{MCT}$ and its higher activation energy is related to the value and 
temperature dependence of the vertex.  This analysis further reveals that the breakdown of MCT at low temperature might be partially related to
 the strong temperature dependence of 
the vertex. 
 
  As mentioned earlier the AG theory which is based on activation dynamics can completely describe the mode coupling theory (MCT) power law behavior
 in the region where the latter is found to be valid. Since the configurational entropy has a finite value at the MCT transition temperature,
 $T_{c}$, the AG relation is not expected to predict any avoided transition in this regime. Our study reveals that although $S_{c}$ is finite, $S_{c2}$ 
vanishes at $T_{K2}$ (where $T_{K2}= T_c$), thus being responsible
 for the divergence like behavior. However we show that the pair configurational entropy although predicts the correct MCT transition temperature it by itself cannot
predict the MCT power law behaviour. The residual multiparticle entropy (RMPE) plays an important role in providing the correct temperature dependence of 
relaxation time. We also obtain  a connection between the AG coefficient (A),
 pair thermodynamic fragility ($K_{T2}$ and 
 MCT critical exponent ($\gamma$)
and found although first two quantities are dependent on density and temperature, their ratio which is related to $\gamma$, is 
density-temperature independent .

            Note that although the absolute value of $\Delta S$ is in the similar range both at high and low temperature regimes, in the high temperature regime
 it plays a minor role in determining the dynamics, whereas its role at 
 low temperature becomes central as we approach the avoided transition.
This small positive value of $\Delta S$  playing an important role in predicting the MCT power law behaviour is similar to the prediction of 
unified theory \cite{sarika_PNAS}.  In the unified theory it was shown that in a certain temperature regime many body activated dynamics plays a hidden but central role in predicting the MCT like
behaviour of the total relaxation time.
  Although apparently the MCT does not depend on the properties of landscape, the saddles in the landscape have been found to disappear at $T_c$ 
\cite{sciortino-saddles-prl,sciortino-saddle,wales_saddle,sciortino-reply}. Here we show that $S_{c2}$
 also vanishes at $T_c$. Thus there may be a connection between pair configurational entropy and saddles. It will be also interesting to understand the independent role
of pair configurational entropy and RMPE in the landscape picture. These are important open questions to be addressed in the future work.

  \section{Acknowledgements}
This work has been supported by the Department of Science and Technology (DST), India and CSIR-Multi-Scale Simulation and Modeling project.
 MKN thanks UGC and AB thanks DST for fellowship. Authors thank Prof. Kunimasa Miyazaki for discussions.

\clearpage

\clearpage

\begin{thebibliography}{53}
\makeatletter
\providecommand \@ifxundefined [1]{%
 \@ifx{#1\undefined}
}%
\providecommand \@ifnum [1]{%
 \ifnum #1\expandafter \@firstoftwo
 \else \expandafter \@secondoftwo
 \fi
}%
\providecommand \@ifx [1]{%
 \ifx #1\expandafter \@firstoftwo
 \else \expandafter \@secondoftwo
 \fi
}%
\providecommand \natexlab [1]{#1}%
\providecommand \enquote  [1]{``#1''}%
\providecommand \bibnamefont  [1]{#1}%
\providecommand \bibfnamefont [1]{#1}%
\providecommand \citenamefont [1]{#1}%
\providecommand \href@noop [0]{\@secondoftwo}%
\providecommand \href [0]{\begingroup \@sanitize@url \@href}%
\providecommand \@href[1]{\@@startlink{#1}\@@href}%
\providecommand \@@href[1]{\endgroup#1\@@endlink}%
\providecommand \@sanitize@url [0]{\catcode `\\12\catcode `\$12\catcode
  `\&12\catcode `\#12\catcode `\^12\catcode `\_12\catcode `\%12\relax}%
\providecommand \@@startlink[1]{}%
\providecommand \@@endlink[0]{}%
\providecommand \url  [0]{\begingroup\@sanitize@url \@url }%
\providecommand \@url [1]{\endgroup\@href {#1}{\urlprefix }}%
\providecommand \urlprefix  [0]{URL }%
\providecommand \Eprint [0]{\href }%
\providecommand \doibase [0]{http://dx.doi.org/}%
\providecommand \selectlanguage [0]{\@gobble}%
\providecommand \bibinfo  [0]{\@secondoftwo}%
\providecommand \bibfield  [0]{\@secondoftwo}%
\providecommand \translation [1]{[#1]}%
\providecommand \BibitemOpen [0]{}%
\providecommand \bibitemStop [0]{}%
\providecommand \bibitemNoStop [0]{.\EOS\space}%
\providecommand \EOS [0]{\spacefactor3000\relax}%
\providecommand \BibitemShut  [1]{\csname bibitem#1\endcsname}%
\let\auto@bib@innerbib\@empty

\bibitem{sjogren}
{\sc W.~G{\"o}tze} and {\sc L.~Sj{\"o}gren},
\newblock {\em Zeitschrift f{\"u}r Physik B Condensed Matter} {\bf 65}, 415
  (1987).

\bibitem{Gotze}
{\sc W.~G{\"o}tze},
\newblock {\em Journal of Physics: Condensed Matter} {\bf 11}, A1 (1999).

\bibitem{tarjus_prl}
{\sc L.~Berthier} and {\sc G.~Tarjus},
\newblock {\em Phys. Rev. Lett.} {\bf 103}, 170601 (2009).

\bibitem{tarjus_pre}
{\sc L.~Berthier} and {\sc G.~Tarjus},
\newblock {\em Phys. Rev. E} {\bf 82}, 031502 (2010).

\bibitem{tarjus_epje}
{\sc L.~Berthier} and {\sc G.~Tarjus},
\newblock {\em EPJE} {\bf 34}, 96 (2011).

\bibitem{tarjus_berthier_jcp}
{\sc L.~Berthier} and {\sc G.~Tarjus},
\newblock {\em J. Chem. Phys.} {\bf 134}, 214503 (2011).

\bibitem{coslovich-jcp}
{\sc D.~Coslovich},
\newblock {\em J. Chem. Phys.} {\bf 138}, 12A539 (2013).

\bibitem{coslovich}
{\sc D.~Coslovich},
\newblock {\em Phys. Rev. E} {\bf 83}, 051505 (2011).

\bibitem{tarjus}
{\sc D.~Kivelson}, {\sc S.~A. Kivelson}, {\sc X.~Zhao}, {\sc Z.~Nussinov}, and
  {\sc G.~Tarjus},
\newblock {\em Physica A: Statistical Mechanics and its Applications} {\bf
  219}, 27  (1995).

\bibitem{Smarajit_pre}
{\sc S.~Karmakar} and {\sc I.~Procaccia},
\newblock {\em Phys. Rev. E} {\bf 86}, 061502 (2012).

\bibitem{Paddy-jcp13}
{\sc A.~Malins} and {\sc et~al.},
\newblock {\em J. Chem. Phys.} {\bf 138}, 12A535 (2013).

\bibitem{hocky-prl}
{\sc G.~M. Hocky}, {\sc T.~E. Markland}, and {\sc D.~R. Reichman},
\newblock {\em Phys. Rev. Lett.} {\bf 108}, 225506 (2012).

\bibitem{bssb}
{\sc A.~Banerjee}, {\sc S.~Sengupta}, {\sc S.~Sastry}, and {\sc S.~M.
  Bhattacharyya},
\newblock {\em Phys. Rev. Lett.} {\bf 113}, 225701 (2014).

\bibitem{Rosenfeld}
{\sc Y.~Rosenfeld},
\newblock {\em Phys. Rev. E} {\bf 62}, 7524 (2000).

\bibitem{adam-gibbs}
{\sc G.~Adam} and {\sc J.~H. Gibbs},
\newblock {\em J. Chem. Phys.} {\bf 43}, 139 (1965).

\bibitem{Rosenfeld-iop}
{\sc Y.~Rosenfeld},
\newblock {\em J Phys: Condens. Matter} {\bf 11}, 5415 (1999).

\bibitem{Baranyai-cp}
{\sc I.~Borzsák} and {\sc A.~Baranyai},
\newblock {\em Chem. Phys.} {\bf 165}, 227  (1992).

\bibitem{Dzugutov}
{\sc M.~Dzugutov},
\newblock {\em Nature} {\bf 381}, 6578 (1996).

\bibitem{hoyt}
{\sc J.~J. Hoyt}, {\sc M.~Asta}, and {\sc B.~Sadigh},
\newblock {\em Phys. Rev. Lett.} {\bf 85}, 594 (2000).

\bibitem{trusket_2006}
{\sc J.~Mittal}, {\sc J.~R. Errington}, and {\sc T.~M. Truskett},
\newblock {\em J. Chem. Phys.} {\bf 125}, 076102 (2006).

\bibitem{Ruchi_charu_2006}
{\sc R.~Sharma}, {\sc S.~N. Chakraborty}, and {\sc C.~Chakravarty},
\newblock {\em J. Chem. Phys.} {\bf 125}, 204501 (2006).

\bibitem{zwanzig}
{\sc R.~Zwanzig},
\newblock {\em PNAS} {\bf 85}, 2029 (1988).

\bibitem{saikat_jcp}
{\sc S.~Banerjee}, {\sc R.~Biswas}, {\sc K.~Seki}, and {\sc B.~Bagchi},
\newblock {\em J. Chem. Phys.} {\bf 141}, 124105 (2014).

\bibitem{alok}
{\sc A.~Samanta}, {\sc S.~M. Ali}, and {\sc S.~K. Ghosh},
\newblock {\em Phys. Rev. Lett.} {\bf 87}, 245901 (2001).

\bibitem{shankar_das}
{\sc C.~Kaur}, {\sc U.~Harbola}, and {\sc S.~P. Das},
\newblock {\em J. Chem. Phys.} {\bf 123}, 034501 (2005).

\bibitem{manish_charu}
{\sc M.~Agarwal}, {\sc M.~Singh}, {\sc B.~Shadrack~Jabes}, and {\sc
  C.~Chakravarty},
\newblock {\em J. Chem. Phys.} {\bf 134}, 014502 (2011).

\bibitem{shila-jcp}
{\sc Sengupta}, {\sc Shiladitya}, {\sc F.~Vasconcelos}, {\sc F.~Affouard}, and
  {\sc S.~Sastry},
\newblock {\em J. Chem. Phys.} {\bf 135}, 194503 (2011).

\bibitem{shila-prl}
{\sc S.~Sengupta}, {\sc S.~Karmakar}, {\sc C.~Dasgupta}, and {\sc S.~Sastry},
\newblock {\em Phys. Rev. Lett.} {\bf 109}, 095705 (2012).

\bibitem{sciortino-pre-2002}
{\sc S.~Mossa}, {\sc E.~La~Nave}, {\sc H.~E. Stanley}, {\sc C.~Donati}, {\sc
  F.~Sciortino}, and {\sc P.~Tartaglia},
\newblock {\em Phys. Rev. E} {\bf 65}, 041205 (2002).

\bibitem{sciortino-prl-2000}
{\sc E.~La~Nave}, {\sc A.~Scala}, {\sc F.~W. Starr}, {\sc F.~Sciortino}, and
  {\sc H.~E. Stanley},
\newblock {\em Phys. Rev. Lett.} {\bf 84}, 4605 (2000).

\bibitem{sciortino-prl-2002}
{\sc E.~La~Nave}, {\sc H.~E. Stanley}, and {\sc F.~Sciortino},
\newblock {\em Phys. Rev. Lett.} {\bf 88}, 035501 (2002).

\bibitem{kob}
{\sc W.~Kob} and {\sc H.~C. Andersen},
\newblock {\em Phys. Rev. E} {\bf 51}, 4626 (1995).

\bibitem{chandler}
{\sc J.~D. Weeks}, {\sc D.~Chandler}, and {\sc H.~C. Andersen},
\newblock {\em J. Chem. Phys.} {\bf 54}, 5237 (1971).

\bibitem{lammps}
{\sc S.~J. Plimpton},
\newblock {\em J. Comput. Phys.} {\bf \textbf{117}}, 1 (1995).

\bibitem{kob-nauroth-pre}
{\sc M.~Nauroth} and {\sc W.~Kob},
\newblock {\em Phys. Rev. E} {\bf 55}, 657 (1997).

\bibitem{srikanth_PRL}
{\sc S.~Sastry},
\newblock {\em Phys. Rev. Lett.} {\bf 85}, 590 (2000).

\bibitem{Srikanth_nature}
{\sc S.~Sastry},
\newblock {\em Nature} {\bf 409}, 164 (2001).

\bibitem{Kirkwood}
{\sc J.~G. Kirkwood} and {\sc E.~M. Boggs},
\newblock {\em J. Chem. Phys.} {\bf 10}, 394 (1942).

\bibitem{green_jcp}
{\sc R.~E. Nettleton} and {\sc M.~S. Green},
\newblock {\em J. Chem. Phys.} {\bf 29}, 1365 (1958).

\bibitem{raveche}
{\sc H.~J. Ravech{\'e}},
\newblock {\em J. Chem. Phys.} {\bf 55}, 2242 (1971).

\bibitem{Wallace}
{\sc D.~C. Wallace},
\newblock {\em J. Chem. Phys.} {\bf 87}, 2282 (1987).

\bibitem{Reichman}
{\sc Y.~Brumer} and {\sc D.~R. Reichman},
\newblock {\em Phys. Rev. E} {\bf 69}, 041202 (2004).

\bibitem{szamel-pre}
{\sc E.~Flenner} and {\sc G.~Szamel},
\newblock {\em Phys. Rev. E} {\bf 72}, 031508 (2005).

\bibitem{manu_under_prep}
{\sc manuscript~under preparation}.

\bibitem{Bengtzelius}
{\sc W.~G. U~Bengtzelius} and {\sc A.~Sjolande},
\newblock {\em J. Phys. C: Solid State Phys} {\bf 17}, 5915 (1984).

\bibitem{leutheusser}
{\sc E.~Leutheusser},
\newblock {\em Phys. Rev. A} {\bf 29}, 2765 (1984).

\bibitem{jansen-reichman}
{\sc L.~M.~C. Janssen}, {\sc P.~Mayer}, and {\sc D.~R. Reichman},
\newblock {\em Phys. Rev. E} {\bf 90}, 052306 (2014).

\bibitem{szamel-gmct}
{\sc G.~Szamel},
\newblock {\em Phys. Rev. Lett.} {\bf 90}, 228301 (2003).

\bibitem{sarika_PNAS}
{\sc S.~M. Bhattacharyya}, {\sc B.~Bagchi}, and {\sc P.~G. Wolynes},
\newblock {\em PNAS} {\bf 105}, 16077 (2008).

\bibitem{sciortino-saddles-prl}
{\sc L.~Angelani}, {\sc R.~Di~Leonardo}, {\sc G.~Ruocco}, {\sc A.~Scala}, and
  {\sc F.~Sciortino},
\newblock {\em Phys. Rev. Lett.} {\bf 85}, 5356 (2000).

\bibitem{sciortino-saddle}
{\sc L.~Angelani}, {\sc R.~Di~Leonardo}, {\sc G.~Ruocco}, {\sc A.~Scala}, and
  {\sc F.~Sciortino},
\newblock {\em J. Chem. Phys.} {\bf 116}, 10297 (2002).

\bibitem{wales_saddle}
{\sc J.~P.~K. Doye} and {\sc D.~J. Wales},
\newblock {\em J. Chem. Phys.} {\bf 118}, 5263 (2003).

\bibitem{sciortino-reply}
{\sc L.~Angelani}, {\sc R.~Di~Leonardo}, {\sc G.~Ruocco}, {\sc A.~Scala}, and
  {\sc F.~Sciortino},
\newblock {\em J. Chem. Phys.} {\bf 118}, 5265 (2003).

\end{thebibliography}
\end{document}